\documentclass{new_tlp}
\usepackage[mathscr]{eucal}
\usepackage{mathptmx}

\newcommand\bcmdtab{\noindent\bgroup\tabcolsep=0pt%
  \begin{tabular}{@{}p{10pc}@{}p{20pc}@{}}}
\newcommand\ecmdtab{\end{tabular}\egroup}

\usepackage{amssymb}

\usepackage{proof}
\usepackage{stmaryrd}
\usepackage{latexsym}
\usepackage{color}
\usepackage{ifthen}
\usepackage[utf8]{inputenc} 
\usepackage{graphicx} 
\usepackage{url} 
\usepackage{multirow}
\usepackage{filecontents,catchfile} 

\newtheorem{definition}{Definition}[section]

\newtheorem{lemma}[definition]{Lemma}
\newtheorem{proposition}[definition]{Proposition}
\newtheorem{corollary}[definition]{Corollary}

\newtheorem{example}{Example}

\newcommand{\myurl}[1]{{\fontsize{9}{9}\url{#1}}}

\newcommand{\mysf}[1]{{\fontsize{10}{10}\textsf{#1}}}

\newcommand*{\QED}[1]{\vspace{-#1}\hfill{\footnotesize\ensuremath{\blacksquare}}}

\newcommand{\PL}{\mysf{Prolog}}

\newcommand{\BPL}{\mysf{Bousi$\sim$Prolog}}
\newcommand{\BPLa}{\mysf{BPL}} 
\newcommand{\HBPLa}{\mysf{HBPL}} 

\newcommand{\fDES}{\mysf{FuzzyDES}}

\newcommand{\SWIPL}{\mysf{SWI-Prolog}}
\newcommand{\SicsPL}{\mysf{SICStus Prolog}}

  {\vspace{0.5\baselineskip}\small\begin{list}{\labelwidth=0pt\itemindent=0pt}\item}%
  {\end{list}\vspace{0.5\baselineskip}}

\newcommand{\dom}{{\cD}om}

\newfont{\msbm}{msbm10 scaled 1000}

\newcommand{\ol}[1]{\overline{#1}}  

\def\ri{<\!\!\!<}   

\def \tuple#1{\langle #1 \rangle} 

\newcommand{\T}{\mbox{\footnotesize $\triangle$}}

\newcommand{\wmgu}{{\sf\small wmgu}_{\cR}^{\lambda}} 

\newcommand{\revto}{\leftarrow}

\newcommand{\hpsld}[1]{\stackrel{\!\!\! #1 \;}{{\Rightarrow}_{\mbox{\tiny HPSLD}}}}

\newcommand{\hsld}[1]{\stackrel{\!\!\! #1 \;}{{\Rightarrow}_{\mbox{\tiny HSLD}}}}

\newcommand{\hwsld}[1]{\stackrel{\!\!\! #1 \;}{{\Rightarrow}_{\mbox{\tiny HWSLD}}}}

\newcommand{\sld}[1]{\stackrel{\!\!\! #1 \;}{{\Rightarrow}_{\mbox{\tiny SLD}}}}

\newcommand{\EXP}[1]{\stackrel{\!\!\! #1 \;}{{\Rightarrow}_{\mbox{\tiny EXP}}}}

\def\defemb#1#2{\expandafter\def\csname #1\endcsname
                              {\relax\ifmmode #2\else\hbox{$#2$}\fi}}
\defemb{cA}{{\cal A}}
\defemb{cB}{{\cal B}}
\defemb{cC}{{\cal C}}
\defemb{cD}{{\cal D}}
\defemb{cE}{{\cal E}}
\defemb{cF}{{\cal F}}
\defemb{cG}{{\cal G}}
\defemb{cH}{{\cal H}}
\defemb{cI}{{\cal I}}
\defemb{cJ}{{\cal J}}
\defemb{cK}{{\cal K}}
\defemb{cL}{{\cal L}}
\defemb{cM}{{\cal M}}
\defemb{cN}{{\cal N}}
\defemb{cO}{{\cal O}}
\defemb{cP}{{\cal P}}
\defemb{cQ}{{\cal Q}}
\defemb{cR}{{\cal R}}
\defemb{cS}{{\cal S}}
\defemb{cT}{{\cal T}}
\defemb{cU}{{\cal U}}
\defemb{cV}{{\cal V}}
\defemb{cX}{{\cal X}}
\defemb{cY}{{\cal Y}}
\defemb{cZ}{{\cal Z}}

\defemb{wR}{\mbox{\footnotesize $\widehat{\cal R}$}}

\long\def\comment#1{}

\newcommand{\todoFlag}{OFF} 
\newcommand{\version}{arxiv} 

\newcommand{\tplp}[2]{\ifthenelse{\equal{\version}{tplp}}{#1}{#2}}
\newcommand{\supp}[2]{\ifthenelse{\equal{\version}{supp}}{#1}{#2}}
\newcommand{\arxiv}[2]{\ifthenelse{\equal{\version}{arxiv}}{#1}{#2}}

\newcommand{\todo}[1]{\ifthenelse{\equal{\todoFlag}{ON}}{\fcolorbox{red}{yellow}{
			\begin{minipage}{0.9\linewidth}#1\end{minipage}
	}}{}}

\usepackage[normalem]{ulem}
\newcommand{\tachar}[1]{\ifthenelse{\equal{\todoFlag}{ON}}{{\color{red}\sout{#1}}}{}}

\renewcommand{\cR}{\mathscr{R}}
\renewcommand{\cQ}{\mathscr{Q}}
\renewcommand{\cE}{\mathscr{E}}
\renewcommand{\cV}{\mathscr{V}}
\renewcommand{\cD}{\mathscr{D}}
\renewcommand{\cB}{\mathscr{B}}

\supp{
\title[Theory and Practice of Logic Programming]{Appendices for the Paper ``Planning for an Efficient Implementation of Hypothetical Bousi$\sim$Prolog''}
}
{
\title[Theory and Practice of Logic Programming]{Planning for an Efficient Implementation of\\ Hypothetical Bousi$\sim$Prolog}
}

\author[P. JULI\'AN-IRANZO and F. S\'AENZ P\'EREZ]
         {Pascual Juli\'an-Iranzo\thanks{This work was supported by the State Research Agency (AEI) of the Spanish Ministry of Science and Innovation under grant PID2019-104735RB-C42 (SAFER), by the Spanish Ministry of Economy and Competitiveness, under the grants TIN2016-76843-C4-2-R (MERINET), TIN2017-86217-R (CAVI-ART-2), and by the Comunidad de Madrid, under the grant S2018/TCS-4339 (BLOQUES-CM), co-funded by EIE Funds of the European Union.}\\
          Dept. of Information Technologies and Systems,
          University of Castilla-La Mancha, Spain\\
          \email{Pascual.Julian@uclm.es}
          \and Fernando S\'aenz-P\'erez\\
          Dept. of Software Engineering and Artificial Intelligence,
          Universidad Complutense de Madrid, Spain\\
          \email{fernan@sip.ucm.es}}
          
\jdate{March 2003}
\pubyear{2003}
\pagerange{\pageref{firstpage}--\pageref{lastpage}}
\doi{S1471068401001193}

\begin{document}

\maketitle

\supp{
}
{
\begin{abstract}
This paper explores the integration of hypothetical reasoning into an efficient implementation of the fuzzy logic language \textsf{Bousi$\sim$Prolog}.
To this end, we first analyse what would be expected from a logic inference system, equipped with what is called embedded implication, to model solving goals with respect to assumptions.
We start with a propositional system and incrementally build more complex systems and implementations to satisfy the requirements imposed by a system like \textsf{Bousi$\sim$Prolog}.
Finally, we propose an inference system, operational semantics, and the translation function to generate efficient \textsf{Prolog} programs from \textsf{Bousi$\sim$Prolog} programs.
{\color{black}\arxiv{This paper is under consideration for acceptance in TPLP.}{}}
\end{abstract}
}
\supp{ 
}
{
}
\supp{
}
{
\section{Introduction}
\label{sec-intro}

Hypothetical reasoning allows deductions to be made in terms of assumed data.
Applications include planning and scheduling \cite{planningTransport2008}, logistics \cite{jbl.12166}, behaviour \cite{DBLP:conf/ieaaie/BosseG09}, healthcare \cite{DBLP:conf/mobihealth/MinutoloEP16}, law \cite{DBLP:journals/ail/Bench-CaponP10} and everything related to decision making in the scenarios envisioned.
This so-called `what-if' analysis \cite{Rizzi2009} plays an important role in saving resources, time and money.
For example, designing the pipe network of a gas company includes an assessment of the appropriateness of conversion stations and pipes in terms of gas production, sourcing and expected client demand.

In addition, real-world applications must typically handle not-yet-known data, foreseeing scenarios for which it is necessary to handle vague information.
Fuzzy logic is suitable for such imprecise and subjective knowledge, and has been successfully applied to such different fields as process control (e.g., UAV flight control \cite{MV14JSEA}, Japan Sendai Subway with its ATC/ATO security system, domestic appliances such as Samsung dishwashers\ldots), deductive databases (FSQL \cite{galindo2005newFSQL}, a fuzzy version of SQL), and fuzzy expert systems (such as \cite{fuzzyexpertsystem2013} in the healthcare domain).

Combining features from both hypothetical reasoning and fuzzy logic thus seems a reasonable field to study.
This paper deals with such a combination for the inclusion of hypothetical reasoning (with features derived from intuitionistic logics) in the \BPL\  (\BPLa) fuzzy logic system {\color{black} \cite{RJ14JIFS,JR17FSS}}.
In particular, we focus on embedded implications, as in \cite{bonner1994hypothetical}.
Assumptions can be reused throughout proofs in this kind of logic, in contrast to substructural logics \cite{lopez_pimentel:1999:490}.
Basically, the following inference rule, which uses so-called embedded implication `$\Rightarrow$' \cite{bonner1994hypothetical}, captures the fact that hypothesis $R$ (a rule) is used (as many times as needed) to derive the proof for goal $\phi$ in the context of program $\Pi$:
\[\infer[]{\Pi \vdash R \Rightarrow \phi}{\Pi \cup \{R\} \vdash \phi}\]
\noindent This inference rule means that if the inference expression above the line can be inferred, then the one below the line can be inferred too. However, this inference rule, linked to the deduction theorem of classical logic, only models a small part of the characteristics and power of hypothetical reasoning.

This paper explores a path to an efficient implementation of hypothetical reasoning in \textsf{Bousi$\sim$} \textsf{Prolog}.
As developed for \SWIPL\ \cite{wielemaker:2011:tplp}, this \PL\ system is considered here for implementations.
Firstly, Sections \ref{sec-prop-hyp} to \ref{sec-pred-hyp} incrementally develop a hypothetical logic inference system.
Specifically, Sections \ref{sec-hyp-precomp} and \ref{sec-subst-in-assump} introduce the requirements that \BPL\ imposes on hypothetical reasoning that need to be tackled and how they can be addressed in the hypothetical logic inference system.
Section \ref{sec-applying-bpl} recalls the formalisation of \BPL\ and adapts it to efficiently include hypothetical reasoning in this language and system.
Finally, Section \ref{sec-conclusions} gives the conclusions and sets out further work to be addressed. 
{\color{black}
\tplp{Supplementary material corresponding to this paper is provided at the TPLP archives in the form of three appendices which contain a performance analysis, formal proofs and a discussion on related work.}{Three appendices contain a performance analysis, formal proofs and a discussion on related work.}
}

\section{Implementing a hypothetical propositional logic inference system}
\label{sec-prop-hyp}

As a first step, we study how to introduce hypothetical reasoning in a propositional logic framework where a \emph{program} $\Pi$ comprises \emph{rules} of the form $A \leftarrow Q_1 \land \ldots \land Q_n$, where $A$ is a propositional letter, called the \emph{head} of the rule; and $Q_i$ are propositional letters or embedded implications which form the \emph{body}.
A rule $A\leftarrow true$ is written as $A$.
An \emph{embedded implication} is an expression of the form $R_i \Rightarrow A_i$ where $R_i$ is a rule (possibly with embedded implications in its body) and $A_i$ is a propositional letter.  Embedded implications cannot appear as the head of a rule. Note also that embedded implications can be nested. That is,  the expression $R_1 \Rightarrow ( \ldots \Rightarrow (R_n \Rightarrow A_i) \ldots)$ is a well-formed expression of this language.
In the following, we shall write nested embedded implications as $R_1 \Rightarrow \ldots \Rightarrow R_n \Rightarrow A_i$, omitting the parentheses.

\subsection{Hypothetical propositional SLD resolution}
A basic hypothetical propositional logic transition system  
is first defined, to determine the operational semantics of the language.

Let $\Pi$ be a program. Let $E$ be a set of tuples $\tuple{G, \Pi}$ (goal, program), each representing a {\em state} of a computation. Let $\hpsld{} \subseteq (E \times E)$ be the transition relation as now defined in Definition \ref{def-HPSLD-rules}. A {\em successful inference sequence} for a given state is a sequence ending in the state  $\tuple{\square,\Pi}$, where $\square$ represents the empty clause.

\begin{definition}
\label{def-HPSLD-rules}
	{\em Hypothetical Propositional SLD} (HPSLD) {\em resolution} is defined as a transition system $\tuple{E, \hpsld{}}$, 
	and whose transition relation 
	$\hpsld{}$ is the smallest relation  satisfying:\\[1ex]
	{\sf Rule 1:} 
	if $R\equiv(A \leftarrow Q) \in {\Pi}$:
		${\tuple{(\leftarrow \!\!A\!\wedge\!Q'), \Pi}}
		\hpsld{}
		{\tuple{(\leftarrow \!\!Q\!\wedge\!Q'), \Pi}}$
	
	\noindent
	{\sf Rule 2:} 
	if $\tuple{(\leftarrow A), \Pi'} \hpsld{*} \tuple{\square, \Pi'}$, with $\Pi' = \Pi \cup \{R\}$, is a successful inference sequence:
		${\tuple{(\leftarrow \!\!(R\!\Rightarrow\!A)\!\wedge\!Q), \Pi}}
		\hpsld{}
		{\tuple{(\leftarrow \!Q), \Pi}}$	
	\QED{0pt}
\end{definition}

The following simple program, adapted from \cite{bonner1994hypothetical}, illustrates this definition.
\begin{example}
Given the program $\Pi = \{a \leftarrow (d\Rightarrow b)\wedge e, b \leftarrow c, c \leftarrow d, e \}$   
and the goal $\leftarrow a$, the successful derivation 
$ \tuple{\leftarrow a, \Pi} \hpsld{} \tuple{\leftarrow (d\Rightarrow b)\wedge e, \Pi} 
		\hpsld{} \tuple{\leftarrow e, \Pi}  \hpsld{} \tuple{\square, \Pi}$
is possible thanks to the hypothetical query associated derivation: 
$ \tuple{\leftarrow b, \Pi\cup\{d\}} \hpsld{} \tuple{\leftarrow c, \Pi\cup\{d\}}  
	\hpsld{} \tuple{\leftarrow d, \Pi\cup\{d\}} \hpsld{} \tuple{\square, \Pi\cup\{d\}}$
where the initial program $\Pi$ is expanded with the hypothesis $d$. 
\end{example}

This definition of hypothetical propositional resolution can be neatly implemented in \PL\ with a meta-interpreter keeping track of the current program, which can be augmented as needed by the embedded implication.
Proposals of other logic systems \cite{bonner1994hypothetical,lopez_pimentel:1999:490} also suggest this kind of implementation (passing the program as an argument).
The solving of an embedded implication can be implemented in a meta-interpreter with the clause: $solve((R \Rightarrow \phi), \Pi) \leftarrow solve(\phi, [R|\Pi])$ (cf. 
Appendix A). 
But passing the program as an argument in \PL\ requires that its data structure be built in the heap, a costly operation.
Another alternative is to use the dynamic database to assert the program rules.
In contrast to other \PL\ implementations, \SWIPL\ does not differentiate between compiling and asserting, producing a similar code for both operations which, in particular, retains multi-argument and deep indexing. 
In this way, accessing data in dynamic memory is typically faster than in the heap, even using techniques such as Key-Value associations implemented as a balanced binary tree (AVL tree).

A first approach to dealing with the embedded implication is to consult the program as a normal \PL\ program with a definition for the embedded implication with the operator $\Rightarrow$:
$(X \Rightarrow Y) \leftarrow assertz(X) \land call(Y) \land retract(X).$
Thus, it is possible to answer goal $\leftarrow p$ for program \{$p \leftarrow q \Rightarrow q$\}: solving $p$ amounts to asserting $q$ and subsequently solving goal $\leftarrow q$.
When $q$ has been proved, the assumption is retracted.
However, this approach is not correct because alternatives can be lost. 
Consider the program \{$p \leftarrow (q \leftarrow r) \Rightarrow q$, $r$, $r$\}.
Clearly, there should be two answers for goal $p$ but, after solving $q$ for the first answer, the hypothesis is retracted, so that further answers are lost.

Simply removing the retraction does not fix the issue because the hypothesis would then be beyond the scope of the implication.
For example, in the program \{$p \leftarrow (q \Rightarrow q) \land q$\}, the goal $\leftarrow p$ would succeed when it should fail because the second call to $q$ is outside the scope of the embedded implication. 
Instead, a plausible approach is to attach each hypothesis to the program context it belongs to, and allow its selection only in that context, which is the technique explained in the next section.
\subsection{Program contexts to implement embedded implications}
\label{sec-hyp_inf_system}

Rule 2 in Definition \ref{def-HPSLD-rules} constructs a proof for the consequent $A$ in the context of $\Pi \cup \{R\}$, where $\Pi$ is the current program and $R$ is the antecedent of the embedded implication (hypothesis).
Thus, each assumption builds a new program which can be understood as a program context for the proof of a consequent.

\begin{definition} Given a program $\Pi$ and an embedded implication query $R \Rightarrow A$, the \emph{hypothetical context} (or simply a \emph{context}) of a proof for $A$ is the program $\Pi' = \Pi\cup \{R\}$. 
\end{definition}
Because embedded implication queries can be nested, $R_1 \Rightarrow R_2 \Rightarrow \ldots R_n \Rightarrow A$, also hypothetical contexts can be chained $\Pi_0 \subseteq \Pi_1 \subseteq \Pi_2 \subseteq \ldots \subseteq \Pi_n$, where $\Pi_{i+1} = \Pi_i \cup \{R_{i+1}\}$.
Hypothetical contexts  are partially ordered by the set inclusion relation $\subseteq$.
For pragmatic reasons, contexts can be identified by a sequence of unique symbols, one for each new context in the inference sequence. 
For the chain of contexts $\Pi_0 \subseteq \Pi_1 \subseteq \Pi_2 \subseteq \ldots \subseteq \Pi_n$, the sequence of symbols $i_1\ldots i_j$, $1\leq j \leq n$, identifies the context $\Pi_j$, where $\Pi_0$ stands for the initial user program (for which the empty sequence is assumed as its context identifier).

More formally, given a (possibly total ordered) set of indexes $I$, a context $\Pi_{j}$ can be identified by a univocal sequence of indexes $s_j = i_1\ldots i_j$, $1\leq j \leq n$, called a \emph{context identifier}, and alternatively denoted by $\Pi_{s_j}$. For a context $\Pi_s$, with context identifier $s$, and an embedded implication query $R \Rightarrow A$, a new context $\Pi_{s.i}$ is built, where the context identifier $s.i$ is the concatenation of a fresh index $i\in I$ to the sequence $s$. 
The initial user program $\Pi_0$ is denoted by $\Pi_\epsilon$, where $\epsilon$ is the empty sequence.

The set of context identifiers $S$ is a prefix-ordered set, where the prefix relation $\preceq$ on  $S$ is defined  as follows: for two context identifiers $s_1$ and $s_2$, $s_1\preceq s_2$ if and only if $s_1$ is a prefix of $s_2$; that is, there exists a sequence of indexes $s$ such that  $s_2 = s_1.s$. The prefix relation is reflexive ($(\forall s\in S)~ s\preceq s$), anti-symmetric ($(\forall s_1,s_2\in S) s_1\preceq s_2 \wedge s_2\preceq s_1 \to  s_1=s_2$), transitive ($(\forall s_1,s_2,s_3\in S) s_1\preceq s_2 \wedge s_2\preceq s_3 \to  s_1\preceq s_3$) and downward total ($(\forall s_1,s_2,s_3\in S)~ s1\preceq s_3\wedge s_2\preceq s_3 \to (s1\preceq s_2 \vee s_2\preceq s_1)$) \cite{Cui13}.
The first three properties classify a prefix order as a partial order.  Only the downward totality is special, meaning that, although the future evolution of a computation may be branching, from a given point of execution, the past is always totally ordered. In our framework, where the set of sequences has a minimum element (the empty sequence $\epsilon$), the evolution of a computation can be abstracted by a tree structure, where the branches are sequences representing a context identifier and each prefix in the sequence, a previous context identifier.
By construction, for all context identifiers $s_1$ and $s_2$, if $s_1\preceq s_2$ then $\Pi_{s_1} \subseteq \Pi_{s_2}$.

As contexts are sets of rules, each rule can be identified as belonging to a context by the identifier of its program context.
Thus, a context identifier can be  associated with a rule.

\begin{definition}[Scope of a rule]
\label{def-context-inclusion}
Given a rule $R$ with context identifier $s_j$, its \emph{scope} is any context $\Pi_{s_k}$ such that $s_j\preceq s_k$.   \QED{0pt}
\end{definition}

In the framework of hypothetical reasoning, a rule only can be selected for solving if it is within the scope of the current context.
Alternatively, we often say that $R$, with context identifier $s_j$, is \emph{included} in (or \emph{belongs} to) the context $\Pi_{s_k}$, because $\Pi_{s_j} \subseteq \Pi_{s_k}$ when $s_j\preceq s_k$. Thus, from a context $\Pi_{s_k}$ it is possible to access all rules whose context identifier $s_j$ is a prefix of $s_k$.

\subsection{Translating hypothetical propositional programs into Prolog} \label{sec-translating-prop-programs}
Hypothetical propositional SLD resolution can be efficiently implemented by translating hypothetical propositional programs into \PL\ in a such a way as to mimic this operational semantics. 
This translation is based on, first, providing a context identifier for each rule
(either a user-program rule or the hypothesis in an embedded implication);
and, second, transmitting the current context for which the rule is selected during inferencing.
Each hypothesis receives a new index symbol (an integer for simplicity) which is added to the current context identifier sequence to form the identifier of the new context.
This sequence can be implemented as a reversed integer list.

In contrast to the basic implementations in Section \ref{sec-prop-hyp}, user programs and hypotheses must be translated by adding the new arguments for the rule and current context identifiers:

\begin{definition}[Propositional rule and goal translation]  
\label{def-prop-translation}
The {\em rule translation} of $R\equiv p \leftarrow \land_{1\leq i \leq n} q_i$ for a rule context identifier $S^R$ and a current context identifier $S^C$, is defined as
$p(S^R, S^C) \leftarrow chk(S^R, S^C) \land_{1\leq i \leq n} q'_i,$  
where $chk(S^R, S^C)$ checks context inclusion (i.e., whether $S^R \preceq S^C$) and the {\em goal translation} $q'_i$, for the current context identifier $S^{C}$, is either:
\begin{itemize}
\item 
$q_i$ for a built-in call; or 

\item 
$q_i(C, S^{C})$ for a user-predicate call, where $C$ is a new free variable representing any context in the scope of the current context $S^C$; or

\item 
\texttt{$H' \Rightarrow G'$} for an embedded implication $H \Rightarrow G$, 
where:
$H'$ is the rule translation of $H$ for a rule context identifier $S^{H}= S^C.I$ (where $I$ is a fresh index) and a new current context identifier $S^{C'}$, and
$G'$ is the goal translation of $G$ for a current context identifier $S^{H}$.\QED{0pt}
\end{itemize}
\end{definition}

Definition~\ref{def-prop-translation} deserves some comment. Firstly, variables (e.g., $S^{R}$ or $S^{C}$) denote context identifiers, which may be unknown at translation time. In general, $S^{R}$ (or $S^{H}$) denotes the context to which a rule $R$ (or $H$) belongs, while $S^{C}$ (or $S^{C'}$) denotes the current context in which the rule will be selected for solving or the goal will be solved. 

Since an embedded implication $H \Rightarrow G$ generates a new context $S^{H}= S^C.I$, the corresponding hypothetical rule $H$ will be annotated with $S^{H}$, which will be the current context in which the goal G will be solved. It is not possible to set the actual index $I$ in the sequence $S^{H}$ for the embedded implication until solving-time (also, depending on the goal, not all contexts will be needed). 
Actually solving an embedded implication will bind the elements in the sequence representing the context. As $chk(\epsilon, S^C)$ always holds, it can be omitted in the translation. 
The following example illustrates Definition~\ref{def-prop-translation} for a simple program rule:
\begin{example}
Assume the source program rule $R\equiv p \leftarrow q \Rightarrow q$. For a source rule $R$, the rule context identifier $S^{R}=\epsilon$. Applying Definition~\ref{def-prop-translation} to translate $R$ for $S^{R}$, leads to: 
$p(\epsilon, S^C) \leftarrow chk(\epsilon, S^C) \wedge (H' \Rightarrow G')$, 
where $H' \Rightarrow G'$ is the translation of the embedding implication $H \Rightarrow G \equiv q \Rightarrow q$.

In order to translate $H$, consider that $H$ is really seen as the rule $q$ (shorthand for $q\leftarrow true$, where $true$ is a built-in symbol). Then, recursively applying Definition~\ref{def-prop-translation}, the translation of $H$ for the rule context identifier $S^{H}= S^C.I$  
(where $I$ is a fresh variable denoting an unknown index which will be resolved at run-time)
is $H'\equiv q(S^C.I, S^{C'}) \leftarrow chk(S^C.I, S^{C'}) \wedge true$. The translation of $G$ is simply $G'\equiv q(C,  S^C.I)$, where $C$ is a fresh variable.

The translation of $R$ is:
$
p(\epsilon, S^C) \leftarrow  ((q(S^C.I, S^{C'}) \leftarrow chk(S^C.I, S^{C'}) \wedge true) \Rightarrow q(C,  S^C.I))
$
where the call to $chk(\epsilon, S^C)$ has been omitted because, as previously mentioned, it always succeeds.

Our actual implementation produces the following translated \PL\ rule for $R$, where $true$ calls are omitted:
\begin{verbatim}
    p([], A) :- (q([B|A], C) :- chk([B|A], C)) => q(_, [B|A])
\end{verbatim} 
where $chk$ is straightforwardly implemented as: 
\verb+chk(S1, S2) :- append(_, S1, S2)+.
\QED{0pt}
\end{example}

Solving the embedded implication can be implemented as follows:

\begin{definition}[Solving the propositional implication clause]
\label{def-p-impl-solving}
Solving the propositional implication clause for SLD resolution is defined as:\\
$(R \Rightarrow G) \leftarrow
  get\_ci(I) \land
 rule\_context(R, S.I) \land   
  assertz(R) \land
  call(G)$.\\
\noindent where the predicate $get\_ci$/1 returns a unique integer each time it is called, 
and $rule\_context$/2 simply provides access in its second argument to the context of $R$.\QED{0pt}
\end{definition}

Considering $\sld{} \subseteq (E \times E)$ as the transition relation for propositional SLD resolution of a program $\Pi$, where $E$ is a set of states formed by tuples $\tuple{G,\theta,\Pi}$ (goal, substitution, program), the following can be stated:

\begin{proposition}
\label{prop-p-equivalence}
For a program $\Pi$ and goal $\leftarrow A$, there exists $\tuple{(\leftarrow A), \Pi}$ $\hpsld{*} \tuple{\square, \Pi}$ iff $\tuple{\leftarrow A', id, \Pi'} \sld{*} \tuple{\square,id,\Pi'}$, where $\Pi'$ is the propositional translation of each rule in $\Pi$, and $A'$ is the propositional goal translation of $A$. \QED{0pt}
\end{proposition}
See the proof of Proposition~\ref{pred-equivalence} in the  predicate logic case, of which this is a particular case.

\section{Implementing a hypothetical predicate logic inference system}
\label{sec-pred-hyp}

In this section we discuss how to extend hypothetical reasoning from the framework of propositional logic to the framework of predicate logic as a prior step to the introduction of hypothetical reasoning into the fuzzy logic programming language \BPL.

As in the  propositional case, the programs of our hypothetical logic language are sets of \emph{rules} of the form $A \leftarrow Q_1 \land \ldots \land Q_n$, but composed of atomic formulas instead of only propositional letters. Rules are assumed to be universally quantified and the occurrence of extra variables in the body of a rule can be understood as existentially quantified. 
When translating embedded implications, the question of how to deal with their variables is an important one. 

\subsection{Hypothetical  SLD resolution}

Definition \ref{def-HPSLD-rules} can be extended to the non-propositional case by dealing with substitutions, which then become part of the notion of computation state.
Let $E$ be a set of tuples $\tuple{G, \sigma, \Pi}$ (goal, substitution, program), each representing a {\em state} of a computation.
Let $\hsld{} \subseteq (E \times E)$ be the transition relation as defined in Definition \ref{def-HSLD-rules} below.
A {\em successful inference sequence} for a given state is a sequence ending in the state  $\tuple{\square,\sigma',\Pi}$.

\begin{definition}[HSLD resolution]
\label{def-HSLD-rules}
	{\em Hypothetical SLD} (HSLD) {\em resolution} is defined as a transition system $\tuple{E,$ $\hsld{}}$, 
	whose transition relation 
	$\hsld{}$ is the smallest relation  satisfying:\\[1ex]
	{\sf Rule 1:} 
	if $R\equiv(A \leftarrow Q) \in {\Pi}$ and $mgu(A,A')=\sigma$:
		${\tuple{(\leftarrow \!\!A'\!\wedge\!Q'), \theta, \Pi}}
		\hsld{}
		{\tuple{(\leftarrow \!\!Q\!\wedge\!Q')\sigma, \theta\sigma, \Pi}}$
	\vspace{0.75ex}
	\noindent
	{\sf Rule 2:} 
	if $\tuple{(\leftarrow A), id, \Pi'} \hsld{*} \tuple{\square, \sigma, \Pi'}$ is
	a successful inference sequence with $\Pi' = \Pi \cup \{R\}$:
		${\tuple{(\leftarrow \!\!(R\!\Rightarrow\!A)\!\wedge\!Q), \theta, \Pi}}
		\hsld{}
		{\tuple{(\leftarrow \!Q\sigma), \theta\sigma, \Pi}}$
\end{definition}

As in the propositional case, HSLD can be implemented as a meta-interpreter, and Definition \ref{def-prop-translation} can easily be extended to the predicate logic case by using predicates instead of propositions:

\begin{definition}[Rule and goal translation for predicate logic]  
\label{def-predicate-translation}
The {\em rule translation} of 
$R\equiv p(\overline{X}) \leftarrow \land_{1\leq i \leq n} q_i(\overline{X_i})$ for a rule context identifier $S^R$ and a current context identifier $S^C$, 
is defined as the rule
$p(\overline{X},S^R, S^C) \leftarrow chk(S^R, S^C) \land_{1\leq i \leq n} q'_i,$  
where $chk(S^R, S^C)$ checks context inclusion (i.e., whether $S^R \preceq S^C$) 
and the {\em goal translation} of
$q_i(\overline{X_i})$ for the current context identifier $S^{C}$ is either:
\begin{itemize}
\item 
$q_i(\overline{X_i})$ for a built-in call; or 

\item 
$q_i(\overline{X_i}, C, S^{C})$ for a user-predicate call, where $C$ is a new free variable representing any context within the scope of the current context $S^C$; or

\item 
\texttt{$H' \Rightarrow G'$} for an embedded implication $H \Rightarrow G$, 
where:
$H'$ is the rule translation of $H$ for a rule context identifier $S^{H}= S^C.I$ (where $I$ is a fresh index) and a new current context identifier $S^{C'}$, and
$G'$ is the goal translation of $G$ for a current context identifier $S^{H}$.\QED{0pt}
\end{itemize}
\end{definition}

The following example illustrates Definition~\ref{def-predicate-translation} for a simple program that will be used in several places hereafter:
\begin{example}\label{ex-predicate-translation}
Given the source program $\Pi = \{R_1: p(X) \leftarrow g(X) \land (q(X) \Rightarrow r), R_2: g(1), R_3: r\leftarrow q(2)\}$, note that  all rules are translated for the rule context identifier $S^{R}=\epsilon$. Specifically,  applying Definition~\ref{def-predicate-translation} to translate $R_1$ for $S^{R}$, gives: 
$p(X, \epsilon, S^C) \leftarrow chk(\epsilon, S^C) \wedge G'_1 \wedge (H'_2 \Rightarrow G'_2)$,  
where $G'_1$ is the goal translation of $G_1 \equiv g(X)$ and $H'_2 \Rightarrow G'_2$ is the translation of the embedding implication $H_2 \Rightarrow G_2 \equiv (q(X) \Rightarrow r)$.

Because $G_1 \equiv g(X)$ is a user-defined predicate,  Definition~\ref{def-predicate-translation} makes $G'_1= g(X, C, S^{C})$. 
On the other hand, since $H_2 \equiv q(X)\leftarrow true$, on recursively applying Definition~\ref{def-predicate-translation}, the translation of $H_2$ for the rule context identifier $S^{H}= S^C.I$  (where $I$ is a fresh variable denoting an unknown index which will be resolved at run-time) is $H'_2\equiv q(X, S^C.I, S^{C'}) \leftarrow chk(S^C.I, S^{C'}) \wedge true$. The translation of $G_2\equiv r$ for rule context identifier $S^{H}$ is simply $G'_2\equiv r(C',  S^C.I)$, where $C'$ is a fresh variable.
Finally, the translation of $R_1$ is:
$$
p(X, \epsilon, S^C) \leftarrow  g(X,C,S^{C}) \wedge ((q(X, S^C.I, S^{C'}) \leftarrow chk(S^C.I, S^{C'}) \wedge true) \Rightarrow r(C',  S^C.I))
$$ 
where the call to $chk(\epsilon, S^C)$ has been omitted because, as previously mentioned, it always succeeds.

It is easy to determine that the translation of $R_2$  is $g(1,\epsilon, S^C)\leftarrow true$ and the translation of $R_3$ is $r(\epsilon, S^C)\leftarrow q(2, C, S^{C})$.

Our implementation produces the following translated \PL\ program for the source program $\Pi$:
\begin{verbatim}
     p(A,[],B) :- g(A,_,B), ((q(A,[C|B],D):-chk([C|B],D)) => r(_,[C|B])).
     g(1,[],_).
     r([],A) :- q(2,_,A).
\end{verbatim}
\end{example}
\QED{20pt}

\vspace{10pt}
Despite the two solutions above for solving the embedded implication by means of a meta-interpreter and program translation, 
our goal is to add this kind of implication to \BPL, which requires complex rule transformations. 

This raises two problems: First, in contrast to the solution in Definition \ref{def-predicate-translation}, 
a more involved translation is required before asserting hypotheses.
This first issue suggests avoiding program compilations at run-time by translating hypotheses at compile-time, 
but this leads to the second problem:
When asserting a rule, all  variables become universally quantified by default, but this is no longer true in the context of an embedded implication. Thus, it will be necessary to take account of substitutions in assumptions.
Taking the logic system shown in Section~\ref{sec-translating-prop-programs} as a starting point, the following sections will cover these two issues.

\subsection{Hypothesis precompilation} \label{sec-hyp-precomp}

One possible approach to addressing the first problem is to translate and assert clauses at compile-time, so that a program does not need any translation at run-time.
Although this involves all hypotheses in the program being translated and asserted, it is highly likely that the memory thus speculatively used  will return a pay-off in terms of run-time gains.

Since it is not known how many times a given embedded implication will be selected for solving, some sort of registration is needed at run-time.
First, the hypothesis can be translated and asserted at run-time and, whenever the embedded implication is selected for solving at run-time, the corresponding hypothesis is registered for the current context.
Each registration will correspond to one alternative of the hypothesis and can be represented by an entry in a new dynamic predicate $reg$/2: one argument for the rule reference and another for the context in which it was registered.

Therefore, registering a rule $R$ for a new context needs a univocal reference $I^R$ to the rule.
A new argument must thus be added to any predicate containing its reference, which 
can be achieved with a predicate similar to $get\_ci$ (Definition \ref{def-p-impl-solving}). 
Definition \ref{def-prop-translation} must be modified to add this argument to head clauses and to translate the embedded implication with the reference to the translated hypothesis.
In addition to the call to the $chk$/2 predicate, each body rule must include a call to $reg$/2 to retrieve, by backtracking, as many alternatives of the hypothesis as were registered in the course of the solving process.

In turn, solving the embedded implication needs to perform this registration of the hypothesis for the current context $S^C$ augmented with this hypothesis.
Then, the embedded implication is added, with two new arguments: the new element in the context identifier sequence $I^C$ and the current context $S^C$, meaning it is no longer a binary operator:
$\Rightarrow(I^R,G,I^C,S^C) \leftarrow get\_ci(I^C) \land assertz(reg(I^R, S^C.I^C)) \land call(G)$

The following examples informally illustrate the proposed translation (which will be formalised in Definition~\ref{def-translation}) and the behaviour of the translated program. 
From here on, context identifiers will be represented as reversed lists of unique indexes.
\begin{example} 
The program \{$p \leftarrow q \Rightarrow q$\} in Section \ref{sec-prop-hyp} would be translated as:\\
$\Pi  = \{  p(1, [~], S^C) \leftarrow~\Rightarrow (0, q(\_, \_, [I^C|S^C]), I^C, S^C), 
q(0, [\_|\_], S^C) \leftarrow~reg(0,C^R) \land chk(C^R,S^C) \}$.

In contrast to the method proposed in Section~\ref{sec-translating-prop-programs}, the hypothesis $q$ of the embedded implication is compiled into a new rule. Three new arguments are added to the head of the rules: a unique rule index, the rule context identifier, and the current context identifier in which the rule will be used at run-time.

Solving the rule defining $p/3$ will call the operator $\Rightarrow$, which will generate a unique identifier $I^C$ and will register the rule defining $q/3$ (the hypothesis) for the current context $[I^C|S^C]$ before submitting the goal $q(\_, \_, [I^C|S^C])$.

Solving by SLD resolution requires the program to be added as a third argument to the state tuple in the transition relation $\sld{}$, since it can be augmented by $assertz$/1.
\end{example}

\begin{example} 
For the program \{$p \leftarrow (q \Rightarrow q \Rightarrow q)$\}, the goal $\leftarrow p$ succeeds with two solutions.
If the first (resp. second) context receives the identifier 0 (resp. 1), the program is translated as:

\[
\begin{array}{ll}
\Pi = & \{p(2, nil, S^C) \leftarrow 
        \Rightarrow([1],
           \Rightarrow([0],
              q(\_, \_, [I^C_{0}, I^C_{1}|S^C]),
              I^C_{0},
              [I^C_{1}|S^C]),
           I^C_{1},
           A),\\
~~ &   q(0, [\_, \_|\_], S^C) \leftarrow
        reg(0, C^R) \land chk(C^R, S^C), 
       q(1, [\_|\_], S^C) \leftarrow
        reg(1, C^R) \land chk(C^R, S^C)\}
\end{array}
\]

\noindent where the facts: 
$reg(1, [0])$ and $reg(0, [1, 0])$
are added while solving the embedded implications.
Thus, the consequent $q(\_, \_, [I^C_{0}, I^C_{1}|S^C])$ can be solved twice at context $[1,0]$ because both hypotheses match the goal at their respective contexts $([1,0]$ and $[0])$.
\end{example}

An advantage of this approach with respect to that given in Section \ref{sec-translating-prop-programs} is that assumptions via recursion require only one translated rule in the memory, rather than needing as many translated rules as assumptions.
The following program illustrates this:

$\Pi = \{ p(0), p(N) \leftarrow N>0 \land N_1~ is~ N-1 \land (q \Rightarrow p(N_1)) \}$

There is only one translated rule for $q$ (the same as in the previous example) and as many entries in $reg$/2 as $N$ in the goal $\leftarrow p(N)$.

\subsection{Handling substitutions in assumptions} \label{sec-subst-in-assump}

An assertion with free variables is a problem in this approach because it is translated at compile-time and the actual substitution is not known before solving-time.

Compared to the approach in Section \ref{sec-translating-prop-programs} (and extended to the predicate logic case in Definition \ref{def-predicate-translation}), 
where each assumption is asserted at run-time, now an assumption is asserted at compile-time without any trace of its actual substitution, which can lead to losing bindings for the shared variables between the hypothesis and its rule. 
The following program illustrates this issue:

\begin{example}
\label{ex-subs-problem-1}
Consider again the program $\Pi = \{p(X) \leftarrow g(X) \land (q(X) \Rightarrow r), g(1), r\leftarrow q(2)\}$ of Example~\ref{ex-predicate-translation}.  For this program, 
the goal $\leftarrow p(X)$ succeeds because the translation of $q$/1 is $q(X, 0, [\_|\_], S^C) \leftarrow
reg(0, C^R) \land chk(C^R, S^C)$, and the call to $r$ succeeds because $q$/1 is registered and the translation of $q(2)$ unifies with the assertion, but it should not (as happens correctly following Definition \ref{def-predicate-translation}).
\QED{0pt}
\end{example}

Moreover, variable sharing is also a problem.
Let us consider the following example:

\begin{example}
\label{ex-subs-problem-2}
In the program
$\Pi = \{p \leftarrow g(X,Y) \land (q(X,Y) \Rightarrow q(1,2)),$ $g(X,X)
\}$
the goal $\leftarrow p$ succeeds when it should not.\QED{0pt}
\end{example}

Substitutions must therefore be passed when solving an embedded implication by using its hypothesis:
Each rule will contain an extra argument containing the list of shared variables.
In addition, the list of shared variables will be added to the rule registration in the context of the embedded implication so that, at solving-time, the translated implication can map the current substitution to these variables.
Finally, each rule body in the translation will include a call to the $reg$ predicate extended with an argument containing these shared variables, making it possible to transmit variable bindings at run-time.
This is formalised in the next subsection.

\subsection{Putting it all together}
\label{sec-hsld-translation}

Satisfying the requirements set out above, we define rule registration as follows:
 
\begin{definition}[Rule registration]
\label{def-rule-registration}
Rule registration is defined as:
$reg\_rule(I^R, [\overline{X^S}], I^C, S^C) \leftarrow assertz(reg(I^R, [\overline{X^S}],  S^C.I^C))$.\QED{0pt}
\end{definition}

Then, the implementation of embedded implication solving, as introduced for the propositional case in Definition \ref{def-p-impl-solving}, is adapted as follows:

\begin{definition}[Solving implication clause]
\label{def-impl-solving}
The solving implication clause for SLD resolution is defined as:\\
$\Rightarrow (I^R, [\overline{X^S}], G, I^C, S^C) \leftarrow
  get\_ci(I^C) \land
  reg\_rule(I^R, [\overline{X^S}], I^C, S^C) \land
  call(G)$.\QED{0pt}
\end{definition}

Considering what must be added to the translation (precompiling with registering, and variable sharing), rule and goal translation are defined  as follows:

\begin{definition}[Rule and goal translation]
\label{def-translation}
Given a program $\Pi$, the {\em rule translation} of 
$R\equiv p(\overline{s}) \leftarrow \cQ$ in $\Pi$,
for the rule context identifier $\epsilon$ and  
an empty list of variables, is the set of rules consisting of the rule: 
\begin{itemize}
\item $p(\overline{s},  [~], I^R, \epsilon, S^C) \leftarrow  \cQ'$   
where  
$I^R$ is a fresh rule identifier, $S^{C}$ a variable that will be bound to a current context identifier, and  $\cQ'$ is the goal translation of the possibly conjunctive goal $\cQ$.

\item and each rule $H'$ resulting from each embedded implication in $R$ (as defined below).
\end{itemize}

The {\em goal translation} of a goal, $\cQ$, for a current context identifier $S^{C}$, is either:
\begin{itemize}
\item $q_i(\overline{t_i})$ if  $\cQ \equiv q_i(\overline{t_i})$ is a built-in call; or

\item $q_i(\overline{t_i},L,I,C,S^C)$ if  $\cQ \equiv q_i(\overline{t_i})$ is a user predicate call, where $L$ , $I$ and $C$ are new free variables representing any list of shared variables, any index  and any context within the scope of the current context $S^C$ respectively; or

\item ${\cQ'}_1 \land {\cQ'}_2$ if  $\cQ \equiv {\cQ}_1 \land {\cQ}_2$ is a conjunctive goal call, where ${\cQ'}_1$ and ${\cQ'}_2$ are respectively the goal translation of ${\cQ}_1$ and ${\cQ}_2$; or

\item $\Rightarrow (I^{H},[X'^S], G', I^C, S^C)$ if  $\cQ \equiv (H \Rightarrow G)$ is an embedded implication (i.e, $H$ is a rule and $G$ is an atom or an embedded implication), 
where $I^{H}$ is a fresh rule identifier, $[\overline{X'^S}]$ is a list of shared variables between the assumptions in $H$ and the rest of the rule $R$, $I^C$ is a new free variable representing any index, $S^C$ is the current context identifier:
\begin{itemize} 
\item $G'$ is the goal translation of $G$ for a current context identifier $S^{H}= S^C.I^C$, and

\item $H'$ is the rule translation of $H \equiv r(\overline{t}) \leftarrow \cB$, for the rule context identifier $S^H= S^C.I^C$, a current context identifier $S^{C}$, and a list of shared variables $[\overline{X'^S}]$, between the assumptions in $H$ and the rest of the rule  $R$: 
$$
r(\overline{t}, [\overline{X'^S}], I^{H}, S^{H}, S^C) \leftarrow reg(I^H, [\overline{X'^S}], C^R) \land chk(C^R, S^C) \land \cB'
$$ 
where  
$I^H$ is a fresh rule identifier, 
$reg$/3 is the predicate that stores each rule registration identified by $I^H$ for its current context identifier $C^R$ (cf. Definition \ref{def-impl-solving}), 
$chk(C^R, S^C)$ checks whether $C^R \preceq S^C$, and $\cB'$ is the goal translation of $\cB$.
As mentioned above, $H'$ is added to the translated program.   \QED{0pt}
\end{itemize}

\end{itemize}
\end{definition}

\begin{example}
Consider once again the program in Example \ref{ex-predicate-translation}. Following Definition~\ref{def-translation}, the translation of the rule $\{R\equiv p(X) \leftarrow g(X) \land (q(X) \Rightarrow r)$, for the rule context identifier $\epsilon$ and no shared variables, is  
$$p(X,[~],1, \epsilon, S^C) \leftarrow g(X, L, I, C, S^C) \land \Rightarrow(0,[X],r(L', I', C', S^C.I^C),I^C,S^C),$$ 
because 
$g(X, L, I, C, S^C)$ is the translation of the goal $g(X)$ for a current context identifier $S^{C}$ and 
$\Rightarrow(0,[X],r(L', I', C', S^C.I^C),I^C,S^C)$ is the goal translation of $(H \Rightarrow G)\equiv (q(X) \Rightarrow r)$ for $S^{C}$, where rule identifier $I^H$ has been set to the fresh value $0$, the list of variables shared between the assumptions in $H$ and the rest of the rule $R$ is $[X]$, and $r(L', I', C', S^C.I^C)$ is the goal translation of $G\equiv r$ for the current context identifier $S^{H}= S^C.I^C$.

Finally, rule $H\equiv q(X)\leftarrow true$, of the single embedded implication, is translated for the rule context identifier $S^H$, and is added as a new rule (identified by $0$): 
$$q(X,[X],0,S^C.I^C,{S^C}') \leftarrow reg(0, [X], C^R) \land chk(C^R, {S^C}') \land true.$$ 

From Definition~\ref{def-translation}, it is easy to obtain the translation of the remaining rules. 

Our actual implementation produces the following translated \PL\ program for the source program $\Pi$:
\begin{verbatim}
p(A,[],1,[],B) :- g(A, _, _, _, B), (=>(0,[A], r(_, _, _, [C|B]),C, B)).
g(1,[],2,[],_).
r([],3,[],A) :- q(2, _, _, _, A).

q(A,[A],0,[_|_],C) :- reg(0, [A], B), chk(B, C).
\end{verbatim}

It is noteworthy that the last translation fixes the problem mentioned in Example \ref{ex-subs-problem-1}.
\QED{0pt}
\end{example}

Considering $\sld{} \subseteq (E \times E)$ as the transition relation for SLD resolution of a program $\Pi$, where the space state $E$ is a set of tuples $\tuple{G,\theta,\Pi}$ (goal, substitution, program), the following can be stated:

\begin{proposition}
\label{pred-equivalence}
For a program $\Pi$ and a goal $\leftarrow \cQ$, there exists $\tuple{(\leftarrow \cQ), id, \Pi} \hsld{*} \tuple{\square, \sigma, \Pi}$ iff $\tuple{\leftarrow \cQ', id, \Pi'}$ $\sld{*} \tuple{\square,\sigma',\Pi''}$, where   
$\sigma = \sigma'[{\cV}ar(\cQ)]$, 
$\Pi'$ is the translation of each rule in $\Pi$,  $\cQ'$ is the goal translation of $\cQ$, and $\Pi''$ is $\Pi'$ augmented with all the $reg$/3 assertions from embedded implication solving.\QED{0pt}
\end{proposition}

Appendix A
contains a performance analysis and 
Appendix B
the proof of this proposition.

\section{Applying hypothetical reasoning to \BPL}
\label{sec-applying-bpl}

\BPL\ \cite{RJ14JIFS,JR17FSS} is an extension of \PL\ and similarity-based logic programming  \cite{FF99,FF02,LSS01,Ses02}.
We have also developed \fDES\ \cite{JS17,JS18a} as an extension of a Datalog-based deductive database, embedding notions inherited from \BPLa.
We then proposed an extension of \fDES\ to include hypothetical reasoning \cite{DBLP:conf/fuzzIEEE/IranzoS20}, which includes an operational semantics that can be adapted to the hypothetical extension of \BPLa\ (\HBPLa\ for short), as shown below.

\subsection{Formal background}
It is first convenient to review some concepts before extending \BPLa\ to include hypothetical reasoning.

\BPLa, among other features, incorporates a unification algorithm based on proximity relations.   
A proximity relation is a binary fuzzy relation $\cR: U\times U \longrightarrow [0,1]$ on a universe $U$ satisfying, for any $e, e_1,e_2,e_3\in U$, the reflexive ($\cR(e,e)=1$) and symmetric ($\cR(e_1,e_2)=\cR(e_2,e_1)$)) properties. If in addition it has the $\T$-transitive property ($\cR(e_1,e_3)\geq \cR(e_1,e_2)\T\cR(e_2,e_3)$), where the operator $\T$ is an arbitrary t-norm, it is called a {\em similarity} relation.
A proximity relation allows two symbols in a program to be weakly related.
Given a binary fuzzy relation $\cR$ and a value $\lambda$, the $\lambda$-cut  $\cR_{\lambda}=\{\tuple{x,y}\mid \cR(x,y)\geq \lambda\}$. 
By abuse of language, the value $\lambda$ is also called \emph{$\lambda$-cut} (or \emph{cut value}), which can be understood as a user-defined threshold intended to prune answers for a minimum degree of confidence.

A weak unification of terms builds upon the notion of \emph{weak unifier} of level $\lambda$ for 
two expressions $\cE_1$ and $\cE_2$ with respect to $\cR$ (or $\lambda$-unifier): 
a substitution $\theta$ such that $\cR(\cE_1\theta,\cE_2\theta) \geq \lambda$, which is the {\em unification degree} of $\cE_1$ and $\cE_2$ with respect to $\theta$ and $\cR$.
There are several weak unification algorithms  
\cite{JS21}
based on this notion and on the \emph{proximity-based unification relation} $\Rightarrow$, which defines a transition system (based on \cite{MM82}).
This relation, applied to a set of unification problems $\{\cE_i\approx \cE'_i|1\leq i\leq n\}$ can yield either a successful or a failed sequence of transition steps.
In the first case, both a successful substitution and a unification degree are obtained (detailed in, e.g., \cite{JS21}). 
The weak most general unifier (wmgu) $\theta$ between two expressions, denoted by $\wmgu(\cE_1, \cE_2)$, is defined as a $\lambda$-unifier of $\cE_1$ and $\cE_2$ such that there is no other $\lambda$-unifier $\sigma$ which is more general than $\theta$. %
That is, there exists a substitution $\delta$ such that, for any variable $x$ in $\dom(\sigma) \cup \dom(\theta)$,  $\cR(x\sigma, x\theta\delta) \geq \lambda$.
Although, unlike in the classical case, the wmgu is not unique, the weak unification algorithm computes a representative wmgu with approximation degree greater than or equal to any other wmgu.

\subsection{Adapting WSLD resolution for hypothetical reasoning}

To combine WSLD and hypothetical reasoning, it is necessary to augment the notion of computation state by adding to it the context of the computation (i.e., the current program).

Let $E$ be a set of tuples $\tuple{G, \Pi, \theta, \alpha}$ (goal, program, substitution,
approximation degree), each representing a {\em state} of a computation.
Let $\hwsld{} \subseteq (E \times E)$ be the transition relation as defined below in Definition \ref{def-HWSLD}.
A {\em successful inference sequence} for a given state is a sequence ending in the state  $\tuple{\square,\Pi,\theta',\alpha'}$ for some substitution $\theta'$ and approximation degree $\alpha'$, where $\square$ represents an empty clause.

\begin{definition}[HWSLD resolution]\label{def-HWSLD}
	{\em Hypothetical Weak SLD} (HWSLD) {\em resolution} is defined as a transition system $\tuple{E, \hwsld{}}$, 
	whose transition relation 
	$\hwsld{}$ is the smallest relation  satisfying:\\[1ex]
{\sf Rule 1:} 
if $R\equiv\tuple{(A \leftarrow Q);\delta}\ri {\Pi}$,
$\sigma = \wmgu(A, A')\neq fail$, 
$\lambda \leq \beta = \cR(A\sigma,A'\sigma)$, $\cR$ is as defined in $\Pi$, 
and $(\delta \T\beta \T\alpha)\geq \lambda$:
${\tuple{(\leftarrow \!\!A'\!\wedge\!Q'), \Pi, \theta, \alpha}}
    \hwsld{}
    {\tuple{(\leftarrow \!\!Q\!\wedge\!Q')\sigma, \Pi, \theta\sigma, \delta \T\beta \T\alpha}}$
\noindent
{\sf Rule 2:} 
if $\tuple{(\leftarrow A'), \Pi', \theta, \alpha}$ $\hwsld{*} \tuple{\square, \Pi', \sigma, \alpha'}$
is a successful inference sequence:\\
	${\tuple{(\leftarrow \!\!(R\!\Rightarrow\!A')\!\wedge\!Q), \Pi, \theta, \alpha}}
    \hwsld{}
    {\tuple{(\leftarrow \!Q)\sigma, \Pi, \theta\sigma, \alpha\T \alpha'}}$,
where $A$ is an atomic formula, $Q$ and $Q'$ are conjunctions of atomic formulas, $\Pi' = \{R\} \cup \Pi$, and $R\ri\Pi$
indicates that $R$ is a standardised apart rule.\QED{0pt}
\end{definition}

\subsection{Expanded rules: translating hypothetical programs}

A fuzzy logic program $\Pi$ is translated into a logic program by linearising heads, making the weak unification explicit, and explicitly computing the approximation degree. 
Essentially, given a graded rule   
$\tuple{p(\ol{t_n}) \revto Q; \delta}\in\Pi$, 
for each $\cR(p,q)=\alpha$ with $\alpha\geq\lambda$, the following clause is generated:
$$
q(\ol{x_n}) 
\revto (\delta\T\alpha) \wedge x_1\approx t_1 \wedge \dots \wedge x_n\approx t_n \wedge Q
$$
\noindent where $\approx$ is the weak unification operator, $t_i$ are terms, $x_i$ are variables, $\delta$ is the grade of the rule (which may represent the user confidence level in the rule), and $\delta\T\alpha$ abbreviates the goal $\delta\T\alpha\geq\lambda$.
Note that, by reflexivity, $\cR(p,p)=1$ is always in $\cR$.

An operational semantics can be defined for expanded programs by means of a transition system with a set $E$ of states  (goal, program, substitution, degree), adding a new inference rule to the tackling of assumptions:

\begin{definition}[Operational semantics for expanded programs] \label{def-operational-semantics}
The
{\em operational semantics for expanded programs}
is a transition system $\tuple{E,\EXP{}}$
and whose transition relation 
$\EXP{}$ is the smallest relation satisfying: \\[1ex] 
{\sf Rule 1:} if $\beta \in (0,1]$, and $(\beta\T\alpha)\geq\lambda$:
$ 
\tuple{(\leftarrow \underline{\beta}\wedge Q'),\Pi,\theta,\alpha} \: \EXP{}\: 
	 \tuple{(\leftarrow Q'),\Pi,\theta,\beta\T\alpha}
$
\\ \noindent
{\sf Rule 2:} if $\sigma = \wmgu(A,B)\neq fail$, $\lambda \leq \beta = \cR(A\sigma,B\sigma)$, $\cR$ is as defined in $\Pi$, 
and $(\beta\T\alpha)\geq\lambda$:
$
\tuple{(\leftarrow \underline{A\approx B}\wedge Q'),\Pi,\theta,\alpha} \: \EXP{}\: 
	 \tuple{(\leftarrow Q'\sigma),\Pi,\theta\sigma,\beta\T\alpha} 
$
\\ \noindent
{\sf Rule 3:} if $(p(\ol{x_n})\leftarrow\beta\wedge x_1\approx t_1\wedge\dots\wedge x_n\approx t_n\wedge Q)\ri{\Pi}$:
$\tuple{(\leftarrow \underline{p(\ol{s_n})}\wedge Q'),\Pi,\theta,\alpha} 
 \: \EXP{}\: 
 \tuple{(\leftarrow\!\beta\!\wedge\!s_1\!\approx\!t_1\!\wedge\dots\wedge\!s_n\!\approx\!t_n\!\wedge\!Q\!\wedge\! Q'),\Pi,\theta,\alpha} 
$
\\ \noindent
{\sf Rule 4:}
if $\tuple{(\leftarrow p(\ol{s_n})),\Pi',\emptyset,1}  \: \EXP{*}\: \tuple{\square,\Pi',\sigma,\beta}$, 
with $\Pi'=\{R\}\cup\Pi$, is a successful inference sequence:
$
\tuple{(\leftarrow\!\! \underline{R\!\!\Rightarrow\!\!p(\ol{s_n})}\!\wedge\! Q'),\!\Pi,\!\theta,\!\alpha} 
 \!\: \EXP{}\:\! 
 \tuple{(\leftarrow  Q'\sigma),\!\Pi,\!\theta\sigma,\!\beta\T\alpha}  {\scriptstyle}
$ \QED{0pt}
\end{definition}

In this system, transition steps are applied to underlined fragments.

\begin{proposition} \label{prop-correctness}
Given a program $\Pi$, with a proximity relation $\cR$, and 
its expanded program $\Pi'$, there exists a derivation 
$
\tuple{\leftarrow \!\!Q, \Pi, \theta, \alpha} \: \hwsld{*}\: 
	  \tuple{\leftarrow Q', \Pi, \theta', \alpha'}
$
iff there exists a derivation
$
\tuple{\leftarrow \!\!Q, \Pi', \theta, \alpha}$ $\: \EXP{*}\: 
	  \tuple{\leftarrow Q', \Pi', \theta', \alpha'}
$
which computes the same state.\QED{0pt}
\end{proposition}
The proof of this proposition follows a similar structure to the one for Proposition \ref{pred-equivalence}.

\subsection{Implementing the expansion of programs}

The generic implementation of expanded rules, including embedded implications, is done by applying the approach presented in Section \ref{sec-hsld-translation} to the translation of expanded rules in \BPL. 
Thus, instead of working with a program $\Pi$, the following definition refers to program contexts with identifiers $S$ and lists of shared variables $[\overline{X^S}]$.

\begin{definition}[Translation of expanded rules and goals]
\label{def-translation-expanded-programs}
Let $\Pi$ be a \BPLa\ program, $\cR$ a proximity relation on the syntactic domain generated by $\Pi$, $\T$ the fixed t-norm associated with $\cR$, and $\lambda\in [0,1]$ a cut value. Let $R\equiv \tuple{p(\overline{s}) \leftarrow \land_{1\leq i \leq k} q'_i,\delta}$ be a graded rule in $\Pi$. The {\em rule translation} of $R$ for a rule context identifier $\epsilon$, a current context identifier $S^C$ and an empty list of shared variables is a set of expanded rules consisting of:
\begin{itemize}
\item for each $\cR(p,q)=\beta \in \cR$, generate the rule:
\[
\begin{array}{ll}
q(\overline{X}, [], I^R, S^R, S^C,\alpha) \leftarrow &
 over\_\lambda(\beta) \land unify($  $[(X_1,s_1,\alpha_1), \ldots, (X_n,s_n,\alpha_n)])  
\land_{1\leq i \leq k} q'_i \\
~ & \land\  degree\_comp([\delta, \beta,\alpha_1,\ldots, \alpha_n,\alpha'_1, \ldots, \alpha'_{k}], \alpha)
\end{array}
\] 
\item and each rule $H'$ with rule context identifier $S^{H}$ resulting from each embedded implication in $R$ (as defined next), 
\end{itemize}
where  
$I^R$ is a fresh rule identifier, 
$over\_\lambda(\beta)$ checks whether $\beta \geq \lambda$ (i.e., whether the rule can be applied because $\beta$ is over the $\lambda$-cut),  
$unify$/1 computes the unification degrees of the pairs formed by the arguments passed to the variables $\overline{X}$ and the terms $\overline{s}$,
$degree\_comp$/2 compounds the intermediate degrees to obtain the final degree $\alpha$,
and $q'_i$ is the goal translation of $q_i(\overline{t_i})$ for a current context identifier $S^{C}$.

The {\em goal translation} $q'_i $ of $q_i(\overline{t_i})$ for $S^{C}$  is either:
\begin{itemize}
\item $q_i(\overline{t_i})$ for a built-in call; or

\item $q_i(\overline{t_i},L,I,C,S^C, \alpha'_i)$ for a user-predicate call $q_i(\overline{t_i})$, where $L$, $I$, and $C$ are respectively new free variables representing any list of shared variables, any index, and any context within the scope of $S^C$, and $\alpha'_i$ is an approximation degree; or

\item $\Rightarrow (I^{H},[\overline{X'^S}], G', I^C, S^C)$ for an embedded implication $H \Rightarrow G$, 
where $I^{H}$ is a fresh rule identifier,  $[\overline{X'^S}]$ is a list of shared variables between the assumptions in $H$ and the rest of the rule $R$, $I^C$ is a new free variable representing any index, and $S^C$ is the current context identifier:

\begin{itemize}
\item $G'$ is the goal translation of $G$, with degree $\alpha'_i$, for the current context identifier $S^{H}= S^C.I^C$.
\item  Also the rule translation of $H$ for the rule context identifier $S^{H}= S^C.I^C$, a current context identifier $S^C$ and the list $[\overline{X'^S}]$ of shared variables, is generated: 

Let $H\equiv \tuple{r(\overline{s}) \leftarrow \land_{1\leq i \leq l} u_i(\overline{t_i}),\xi}$, 
then the translation of $H$ is a set of expanded rules such that for each $\cR(r,r')=\gamma$,  
the rule H' is generated as: 
\[
\begin{array}{ll}
r'(\overline{X}, [\overline{X^S}], I^R, S^R, S^C,\alpha) \leftarrow  &
over\_\lambda(\xi) \land reg(I^R, [\overline{X^S}], C^R) \land chk(S^R, S^C) \\
~ & \land\ unify([(X_1,s_1,\alpha_1), \ldots, (X_n,s_n,\alpha_n)])  \land_{1\leq i \leq l} u'_i \\
~ & \land\ degree\_comp([\xi, \gamma,\alpha_1,\ldots, \alpha_n,\alpha'_1, \ldots, \alpha'_l], \alpha)
\end{array}
\] 
where 
$reg$/3 is the predicate that stores each rule registration identified by $I^R$ for its current context identifier $C^R$ (cf. Definition \ref{def-rule-registration}), 
$chk(S^R, S^C)$ checks whether $S^R \preceq S^C$,
and $s'_i$ is the goal translation of $s_i(\overline{t_i})$ for a current context identifier $S^{C}$. 
\QED{0pt}
\end{itemize}
\end{itemize}
\end{definition}

\begin{example} \label{ex-HBPL-translation}
Given the \BPL\ program 
$\Pi = \{R_1: \tuple{p(X) \leftarrow g(X) \land (q(X) \Rightarrow r), 0.8}, ~~ 
            R_2: \tuple{g(1), 0.7}, $ 
            $R_3: \tuple{r\leftarrow q(2), 0.9}\}$,  
a proximity relation $\cR$, with an entry $\cR(p,s) = 0.6$, and $\lambda=0.5$, following Definition~\ref{def-translation-expanded-programs}, the translation of $R_1$ for the rule context identifier $\epsilon$ and no shared variables, generates the following set of rules: 
\begin{itemize}
\item For the entry $\cR(p,p)=1$, 
\[
\begin{array}{l}
p(X_1,[~],1,\epsilon,S^C, D) \leftarrow   over\_\lambda(1) \land unify([X_1, X, D_1]) \land g(X, L, I, C, S^C,D_1) \\
~~~~~~~~~~~ \land \Rightarrow(0,[X],r(L', I', C', S^C.I^C,D'_1),I^C,S^C) \land  degree\_comp([0.8, 1, D_1, D'_1], D)
\end{array}
\] 
because 
$g(X, L, I, C, S^C, D_1)$ is the translation of the goal $g(X)$ for a current context identifier $S^{C}$ and 
$\Rightarrow(0,[X],r(L', I', C', S^C.I^C,D'_1),I^C,S^C)$ is the goal translation of $(H \Rightarrow G)\equiv (q(X) \Rightarrow r)$ for $S^{C}$, where rule identifier $I^H$ has been selected to the fresh value $0$, the list of shared variables between the assumptions in $H$ and the rest of the rule $R$ is $[X]$, and $r(L', I', C', S^C.I^C,D'_1)$ is the goal translation of $G\equiv r$ for the current context identifier $S^{H}= S^C.I^C$.

\item For the entry $\cR(p,s)=0.6$, in a similar way,
\[
\begin{array}{l}
s(X_1,[~],2,\epsilon,S^C, D) \leftarrow   over\_\lambda(0.6) \land unify([X_1, X, D_1]) \land g(X, L, I, C, S^C,D_1)  \\
~~~~~~~~~~~ \land \Rightarrow(0,[X],r(L', I', C', S^C.I^C,D'_1),I^C,S^C) \land degree\_comp([0.8, 0.6, D_1, D'_1], D)
\end{array}
\] 
\item Finally, the rule $H\equiv \tuple{q(X)\leftarrow true, 1}$ of the single embedded implication is translated for the rule context identifier $S^H$ and is added as a new rule (identified by $0$): 
\[
\begin{array}{l}
q(X,[X],0,S^C.I^C,{S^C}', D) \leftarrow   over\_\lambda(1) \land reg(0, [X], C^R) \land chk(C^R, {S^C}') \land unify([X_1, X, D_1]) \\
~~~~~~~~~~~  \land true \land degree\_comp([1, 1, D_1], D).$$ 
\end{array}
\] 
\end{itemize}

From Definition~\ref{def-translation-expanded-programs}, it is easy to obtain the translation of the remaining rules:
\begin{itemize}
\item For $R_2$ and the single case in which $\cR(g,g)=1$, 
$g(X_1, [], 3, \epsilon,S^C, D) \leftarrow  over\_\lambda(1) \land unify($ $[X_1, 1, D_1]) \land degree\_comp([0.7, 1, D_1], D)$ 

\item For $R_3$ and  the case $\cR(r,r)=1$, 
$r([], 4, \epsilon,S^C, D) \leftarrow  over\_\lambda(1) \land q(2, L, I,  C, S^C, D).$   
\QED{0pt}
\end{itemize}
\end{example}

For this generic definition, several optimisations are amenable, such as tail-recursion preservation, and deep indexing by linearising only the required head arguments.

\section{Conclusions and Future Work}
\label{sec-conclusions}

This paper has explored the inclusion of hypothetical reasoning in the fuzzy logic language and system \BPL.
We have proposed an operational semantics for hypothetical fuzzy logic programs, and an efficient implementation skeleton, 
which has been assessed in terms of a performance analysis (cf. 
Appendix A) of the proposed technique based on program contexts and precompilation of assumptions.
This analysis confirms the performance benefits in most cases, but it depends on the host \PL\ system: indeed, \SicsPL\ performs better and typically makes the compiled approach behave better than the meta-interpreted one.
In addition, soundness and completeness of the translation are backed by formal results. 

However, incorporation of the approach presented in this paper into the actual system \BPLa\ remains to be done, and thanks to the performance analysis presented in the appendix, we are confident that the final implementation will benefit from such an approach.
Another extension would be to embody negative assumptions, for which an appropriate declarative semantics must be developed before thinking about an operational one.
One may also think of inheriting both positive and negative assumptions of proximity equations, but this is a much more delicate matter performance-wise, because changing relation $\cR$ changes the compilation of the program.
An incremental compilation could perhaps be devised, which would seem to be an appropriate way of tackling this question.
\bibliographystyle{acmtrans}
\bibliography{biblioPaper}
}

\tplp{
}
{
\appendix
\newpage
\section{Performance comparison: meta-interpreted vs. interpreted}
\label{app-performance}

This section analyses the performance of the different solving alternatives explained above: the meta{\color{blue}-}interpreter and the compiler-based  solver interpreter.
They are also compared, when applicable, to the native implementation and the solving in \SWIPL\ 
 of the program under test.
Two alternatives are first described for the meta-interpreter, and then the performance analysis is given, comparing them to the compiled approach in Section 
3. 
The \PL\ interpreter for non-hypothetical programs is also included in the comparison as a reference for the overhead caused by the compiled approach.

\subsection{Hypothetical Meta-interpreters}
Figure \ref{fig-meta-prop} illustrates the hypothetical propositional meta-interpreter, which has been enlarged to deal with disjunctive rules and built-in calls, where a fact is a rule with a $true$ body.
The predicate $builtin$/1 checks whether its argument represents a call to a built-in predicate (different from $\land$, $\lor$ and $\Rightarrow$).

\begin{figure}[h]
$$solve((\phi_1 \land \phi_2), \Pi) \leftarrow solve(\phi_1, \Pi) \land solve(\phi_2, \Pi).$$
$$solve((\phi_1 \lor \phi_2), \Pi) \leftarrow solve(\phi_1, \Pi) \lor solve(\phi_2, \Pi).$$
$$solve(\phi, \Pi) \leftarrow builtin(\phi) \land call(\phi).$$
$$solve((R \Rightarrow \phi), \Pi) \leftarrow solve(\phi, [R|\Pi]).$$
$$solve(\phi, \Pi) \leftarrow member((\phi \leftarrow \phi'), \Pi) \land solve(\phi', \Pi).$$
\caption{Meta-interpreter for hypothetical propositional logic programs}
\label{fig-meta-prop}
\end{figure}

The meta-interpreter for hypothetical propositional logic programs depicted in Figure \ref{fig-meta-prop} is not applicable to predicate logic programs. 
For example, the goal $\leftarrow p$ for the program \{$p \leftarrow q(1)\land q(2), q(X)$\} should succeed, but it does not because solving $\leftarrow q(1)$ creates the substitution \{$X/1$\}, which is not compatible with the second call $\leftarrow q(2)$.
Nonetheless, it can be easily adapted to the non-propositional case by modifying the last clause to:

$$solve(\phi, \Pi) \leftarrow copy\_term(\Pi, \Pi') \land member((\phi \leftarrow \phi'), \Pi') \land solve(\phi', \Pi).$$

However, copying the entire program each time the unification of a rule or fact with the goal is sought, is hugely resource consuming.
A more convenient approach is to look for a matching clause and copy only this clause, as follows:

$$solve(\phi, \Pi) \leftarrow unif\_member((\phi \leftarrow \_), \Pi, R) \land copy\_term(R, (\phi' \leftarrow \phi'')) \land
\phi=\phi' \land solve(\phi'', \Pi).$$

\noindent
where $unif\_member(X,L,Y)$ stands for: $Y$ is a member of $L$ that is unifiable with $X$.

This can be slightly enhanced by using an ordered list for the program predicates (though preserving rule user ordering in each predicate), therefore reducing the serial access time complexity by a factor of 2, on average.
Also, adding cuts for selecting the appropriate meta-interpreter clause will prune choice points in advance, and will also save some tests.
Thus, an actual implementation in \PL\ could be:

\begin{verbatim}
solve((Goal1, Goal2), Program) :-
  !, solve(Goal1, Program), solve(Goal2, Program).
solve((Goal1; Goal2), Program) :-
  !, (solve(Goal1, Program) ; solve(Goal2, Program)).
solve((Hyp => Goal), Program) :-
  !, insert_rule(Hyp, Program, NProgram), solve(Goal, NProgram).
solve(Goal, _Program) :-
  builtin(Goal), !, Goal.  
solve(Goal, Program) :-
  unif_member((Goal :- _Body), Program, (UGoal :- UBody)),
  copy_term((UGoal :- UBody), (CGoal :- CBody)),
  CGoal=Goal, solve(CBody, Program).
\end{verbatim}

Note that, in particular, the last clause will not be selected uselessly, as opposed to the skeleton shown in Figure \ref{fig-meta-prop}.
The predicate \texttt{insert\_rule}/3 inserts a rule in the place corresponding to its ordering.

As suggested by an anonymous referee, there are more alternatives that may be considered.
In particular, this meta-interpreter can be further enhanced by passing only assumed rules to \texttt{solve}/2, instead of the whole program.
Then, the static user program can be accessed by the built-in predicate \texttt{clause}/2.
Moreover, the assumed rules can be better represented by a tree, whose nodes are predicates (key of the tree), with their rules in a list, thereby improving access when many predicates are assumed.
To implement this approach, the last clause of the former meta-interpreter is replaced by:

\begin{verbatim}
% Lookup in the consulted (static) program:
solve(Goal, Program) :-
  clause(Goal, Body), solve(Body, Program).  
% Lookup in the augmented program:
solve(Goal, Program) :-
  functor(Goal, Functor, Arity), atomic_concat(Functor, Arity, Key),
  get_assoc(Key, Program, Rules),
  unif_member((Goal :- _Body), Rules, (UGoal :- UBody)),
  copy_term((UGoal :- UBody), (CGoal :- CBody)),
  CGoal=Goal, solve(CBody, Program).
\end{verbatim}

Association lists are implemented with AVL trees, with $\mathscr{O}(\log{}n)$  worst-case (and expected) time operations, where $n$ denotes the number of elements in the association list \cite{wielemaker:2011:tplp}. 
This meta-interpreter uses \texttt{get\_assoc}/3 in \texttt{solve}/2 to retrieve nodes in the tree and \texttt{put\_assoc}/4 in \texttt{insert\_rule}/3 (cf. its implementation in \texttt{meta2.pl} at the URL mentioned in the next subsection).

\subsection{Performance Analysis}
All experiments were run on an Intel Xeon CPU E3-1505M v5 with 4 physical cores at 2.8 GHz, 16GiB RAM, with the Windows 10 64-bit operating system.
Benchmarks are run on 
the last stable version of \SWIPL\ 64-bit 8.2.4-1 
{\color{black}at the time of writing this}. 
Times in the Tables are given in seconds and are the result of averaging 10 runs (and discarding the first), and individual times were measured with the built-in predicate \texttt{statistics}/2.
We have collected the time for the total run-time measurement (key \textsf{cputime} for \SWIPL), 
which returns the (user) CPU time, and the run-time (key \textsf{runtime}) which returns the total run-time, eliding the time for memory management (garbage collection) and system calls.
Tables include the column \textsf{CPUtime} (for the total run-time), and \textsf{Diftime} (for the total run-time minus the run-time, primarily to reflect the cost of garbage collection).
Garbage collection and stack trimming are carried out before each trial is run; then the time measurement can start.
On completion of the trial, these operations are performed again before taking the elapsed time, in order to account for housekeeping tasks due to running the tests.
Also, \texttt{inferences} is another measurement from  
{\color{black}the statistics predicate,}  
which indicates the number of passes via the call and redo ports in order to execute a goal.
All benchmarks {\color{black} and systems} can be downloaded from 
\url{http://www.fdi.ucm.es/profesor/fernan/iclp2021/Experiments.zip}.

Several classical benchmark programs have been selected (they can be found at 
the \SWIPL\  
site
), 
where some have been adapted to remove cuts.
These programs are intended, firstly, to make clear the price to be paid for including hypothetical reasoning in the system; and, secondly, to compare meta-interpreted alternatives with respect to the compiled alternative.
In any case, it should be recalled that a compiled approach is preferred for the implementation of a system such as \BPLa, because assuming a rule implies a recompilation.
Furthermore, most of those programs have been adapted to build hypothetical versions by assuming either facts or a rule for the data generator.
These versions add a preceding \textit{h} before the classical test name.
With respect to the factorial test program, \textit{facttr} is the tail recursive version, and all benchmark sizes can be consulted at the URL above. 
As a stress test, several parametric hypothetical programs are included, to test embedded implications:
The first (labelled as \textit{hypo1} in the following Tables): \{$p \leftarrow a_1 \Rightarrow a_2 \Rightarrow \ldots \Rightarrow a_n \Rightarrow a_1 \land a_2 \land \ldots \land a_n$\} with a goal $\leftarrow p$ for $n=2000$; 
the second (\textit{hypo2}): \{$p(0) \leftarrow a, p(N) \leftarrow N>0 \land N_1~ is~ N-1 \land (a \Rightarrow p(N_1))$\} with a goal $\leftarrow p(3000)$ and requesting all solutions;
and the third (\textit{hypo3}): \{$p \leftarrow a \Rightarrow \ldots \Rightarrow a \Rightarrow a$\} with a goal $\leftarrow p$ for $3000$ assumptions, also requesting all solutions.
The first program is intended to test the system by assuming a large batch of different predicates iteratively.
The second recursively assumes facts of the same predicate and is intended to analyse the performance in the presence of backtracking by requesting all solutions.
The third is similar to the second but iteratively adding those facts with nested assumptions.

Table 
A 1 
collects the running measurements for classical tests 
and Table 
A 2 
for hypothetical tests.
Rows in the Table include the following labels: 
\textsf{Meta1} for the first meta-interpreter as shown in this appendix,
\textsf{Meta2} for the second, improved, instance,
\textsf{Comp} for the compiled approach as described in Definition 
3.5, and
\textsf{Prolog} for the native execution of (classical, non-hypothetical) tests in 
{\color{black}the} Prolog system.

For classical tests (Table 
A 1
) \textsf{Meta2} performs better than 
\textsf{Meta1} with significant speed-ups, and there is even a notable speed-up in the case of \textit{path}, where recursively traversing the list of the program, including many arcs in the graph, takes a lot of effort that is avoided with \textsf{Meta2}, because the program is consulted. 
Looking now at the compiled alternatives \textsf{Comp} and \textsf{Prolog}, they perform better than the meta-interpreted versions.
Comparing \textsf{Comp} to \textsf{Prolog}, they behave similarly in most cases other than in \textit{nrev} and \textit{queens}, where \textsf{Prolog} is faster {\color{black}(4.62$\times$ and 1.44$\times$, respectively)}.

For hypothetical versions of classical tests (Table 
A 2, 
roughly similar conclusions can be drawn.
Performance of \textsf{Comp} is almost always better than \textsf{Meta2}, and the latter is better than \textsf{Meta1}. 
Only in \textit{hpath} is \textsf{Meta2} faster by a small amount.
Results can be different for other non-classical stress tests such as \textit{hypo1}, for which \textsf{Meta2} performs better. 
In this case, note that AVL trees play an important role in fast accessing of each rule (in \textit{hypo1} there is one assumed rule for each predicate, with a total of 2000). 
In turn, \textsf{Comp} takes more time, both in total time and memory management, even when the number of inferences is roughly a fifth compared to \textsf{Meta1}, but the time taken by memory management is noticeable compared to the other two alternatives. 
{\color{black}This system }\textsf{Comp} performs better for \textit{hypo2} with a similar inference ratio with respect to \textsf{Meta2}. 
With respect to the last test \textit{hypo3}, \textsf{Meta2} is the slowest, while there is a small difference between \textsf{Comp} and the fastest, \textsf{Meta1}.


\CatchFileDef{\test_data_swi}{test_data_swi_classic.tex}{} 

\begin{table}[!h]
\label{tab-swi-classical}
	\scalebox{1}{
		\begin{tabular}{llrrr}
			\hline
			\textsf{Program} & \textsf{System} & \textsf{CPUtime} & \textsf{Diftime} & \textsf{Inferences} \\
			\test_data_swi
       \hline
		\end{tabular}
	}
\caption{
Comparing \textsf{Meta1}, \textsf{Meta2}, \textsf{Comp} and \textsf{Prolog} for classical programs
}
\end{table}

\CatchFileDef{\test_data_swi}{test_data_swi_hypo.tex}{} 

\begin{table}[!h]
\label{tab-swi-hypo}
	\scalebox{1}{
		\begin{tabular}{llrrr}
			\hline
			\textsf{Program} & \textsf{System} & \textsf{CPUtime} & \textsf{Diftime} & \textsf{Inferences} \\

			\test_data_swi
       \hline
		\end{tabular}
	}
\caption{
Comparing \textsf{Meta1}, \textsf{Meta2} and \textsf{Comp} for hypothetical programs
}
\end{table}



%
%
%


\clearpage
\section{Proof for Proposition 
3.6}
\label{app-proofs}

We divide the proof of Proposition 
3.6 into two parts, starting with the proof of the statement ``if there exists a derivation $\cD\equiv (\tuple{(\leftarrow \cQ), id, \Pi} \hsld{*} \tuple{\square, \sigma, \Pi})$ then there exists a derivation $\cD'\equiv (\tuple{\leftarrow \cQ', id, \Pi'} \sld{*} \tuple{\square,\sigma',\Pi''})$''. This direction constitutes a kind of completeness result where we prove that derivations in the original program using the HSLD resolution rule can be reproduced by the SLD operational mechanism in the transformed program. In the second part, we prove the converse of the first statement, that is  ``there exists a derivation $\cD: \tuple{(\leftarrow \cQ), \theta, \Pi} \hsld{*} \tuple{\square, \sigma, \Pi}$ if there exists a derivation $\cD': \tuple{\leftarrow \cQ', \theta, \Pi'} \sld{*} \tuple{\square,\sigma',\Pi''}$'',  which guarantees that the present implementation does not compute answers which are not computed by the HSLD semantics,  leading to a sort of soundness result.

\subsection{Proof of Part I}

\begin{lemma} \label{lem-step-completeness}
Let  $\Pi$ be a program and $\cG\equiv (\leftarrow A\land\cQ_1)$ a goal, if there exists the step 
$$\cS \equiv \tuple{(\leftarrow A\land\cQ_1), \theta, \Pi} \hsld{} \tuple{(\leftarrow \cQ_2\sigma), \theta\sigma, \Pi}$$ 
then there exists the following derivation in the translated program $\Pi'$: 
$$\tuple{(\leftarrow A'\land\cQ'_1), \theta, \Pi'} \sld{+} \tuple{(\leftarrow\cQ'_2(\sigma\cup\delta)),\theta\sigma\cup\delta,\Pi'\cup\Pi_{reg}},$$ 
where $A'$, $\cQ_i'$ are the goal translations of $A$, $\cQ_i$ respectively,
and $\Pi_{reg}$ is the set of all the $reg$/3 assertions due to embedded implication solving. The domain of the substitution $\delta$ shares variables with neither $\theta$ nor $\sigma$, and $\theta\sigma=\theta(\sigma\cup\delta)[{\cV}ar(\cG)]$. \QED{0pt}
\end{lemma}
\begin{proof}
We proceed by induction on the context identifier $s$ associated with the program context.
\begin{enumerate}
\item {\sf Base case($s=\epsilon$)}:  In this case query $A$ must be an atom, that is, $A\equiv p(\overline{s_n})$ and there must be a rule $(p(\overline{t_n}) \leftarrow \cB)\in \Pi$ for which 
$\sigma=mgu(\{p(\overline{s_n})=p(\overline{t_n})\})=mgu(\{\ol{s_n = t_n}\})$ (where $s_i = t_i$, with $1\leq i \leq n$, are unification problems) and step $\cS$ is 
$$(\tuple{\leftarrow p(\overline{s_n})\land\cQ_1, \theta, \Pi} \hsld{} \tuple{\leftarrow (\cB\land\cQ_1)\sigma, \theta\sigma, \Pi})$$  

According to Definition
3.5, the rule translation of $(p(\overline{t_n})  \leftarrow \cB)\in \Pi$ is 
$p(\overline{t_n}, [], 0, \epsilon, S^{C}) \leftarrow \cB'$ and the goal translation of $\leftarrow p(\overline{s_n})$ 
is $\leftarrow p(\overline{s_n}, L, I, C, S)$ where $S^{C}$, $L$, $I$, $C$ and $S$ are fresh variables. It is then easy to verify that:
\[
\begin{array}{l}
  mgu(\{p(\overline{s_n}, L, I, C, S)= p(\overline{t_n}, [], 0, \epsilon, S^{C})\})   \\
  ~~~~~~=mgu(\{\ol{s_n = t_n}, L=[], I=0, C=\epsilon, S=S^{C}\})   \\
  ~~~~~~= \sigma \{L/[], I/0, C/\epsilon, S/S^{C}\} =  \sigma \cup \delta   
\end{array}
\]
where $\delta = \{L/[], I/0, C/\epsilon, S/S^{C}\}$, and its domain does not share variables with $\sigma$. Hence, the following step is possible in the translated program $\Pi'$:
$$\tuple{\leftarrow p(\overline{s_n}, L, I, C, \epsilon)\land\cQ'_1, \theta, \Pi'} \sld{} \tuple{\leftarrow(\cB'\land\cQ'_1)(\sigma \cup \delta), \theta(\sigma \cup \delta), \Pi'}$$
and $\theta\sigma= \theta(\sigma \cup \delta)[{\cV}ar(\cG)]$.

\item {\sf Inductive case ($s > \epsilon$)}: 
The query $A\equiv (H \Rightarrow G)$, that is, an embedded implication (chain), and the step 
$$\cS \equiv \tuple{(\leftarrow (H \Rightarrow G)\land\cQ_1), \theta, \Pi} \hsld{} \tuple{(\leftarrow \cQ_1\sigma), \theta\sigma, \Pi}$$ 
Therefore,  Rule 2 of Definition 
3.1 was applied and the derivation 
 $$\cD_1\equiv (\tuple{(\leftarrow G), id, \Pi_1} \hsld{*} \tuple{\square, \sigma,\Pi_1})$$
must exist, where $\Pi_1 = \Pi\cup\{H\}$ is a new program with context identifier $1>\epsilon$. Then, by the Inductive hypothesis, the following derivation in the translated program $\Pi'$:
$$\cD'_1\equiv (\tuple{(\leftarrow G'), id, \Pi'} \sld{*} \tuple{\square, \sigma\cup\delta,\Pi'\cup\Pi_{reg}})$$
must exist, where $G'$ is the goal translation of $G$, $\Pi_{reg}$ is the set of rule registrations and $\delta$ is a substitution for which its domain shares variables with neither the original (not translated) goal $G$ nor the substitution $\sigma$.

On the other hand, note that the translation of the goal $(H \Rightarrow G)$ 
is $\Rightarrow (0,[\overline{X}], G', I^C, S^C)$ plus the rule translation $H'$ of $H$. 

Now, using Definition 
3.4 for solving embedded implication clauses and Definition 
3.3 of rule registration, and the derivation $\cD'_1$, it is possible to build the following derivation $\cD'$:
{\small
\[
\begin{array}{l}
    \tuple{\leftarrow (\Rightarrow (0, [\overline{X}], G', I^C, S^C)\land\cQ'_1), \theta, \Pi'}  \\
    \sld{} \tuple{\leftarrow (get\_ci(I^C)\land reg\_rule(0, [\overline{X}], I^C, S^C)\land call(G')\land\cQ'_1), \theta, \Pi'}\\
    \sld{} \tuple{\leftarrow (reg\_rule(0, [\overline{X}], 1, S^C)\land call(G')\land\cQ'_1)\{I^C/1\}, \theta\cup\{I^C/1\}, \Pi'}\\
     \sld{} \tuple{\leftarrow (assertz(reg(0, [\overline{X}], S^C))\land call(G')\land\cQ'_1)\{I^C/1\}, \theta\cup\{I^C/1\}, \Pi'} \\   
    \sld{} \tuple{\leftarrow (call(G')\land\cQ'_1)\{I^C/1\}, \theta\cup\{I^C/1\}, \Pi'\cup\{reg(0, [\overline{X}], S^C)\}} \\   
    \sld{} \tuple{\leftarrow (G'\land\cQ'_1)\{I^C/1\}, \theta\cup\{I^C/1\}, \Pi'\cup\{reg(0, [\overline{X}], S^C)\}} \\ 
    \sld{*} \tuple{\leftarrow Q'_1(\sigma\cup\{I^C/1\}\cup\delta), (\theta\sigma)\cup\{I^C/1\}\cup\delta,\Pi'\cup\{reg(0, [\overline{X}], S^C)\}\cup\Pi_{reg}}  
\end{array}
\]
}
where the domain of $\{I^C/1\}\cup\delta$ shares variables with neither $\theta$ nor $\sigma$, and $\theta\sigma= (\theta\sigma\cup\{I^C/1\}\cup\delta)[{\cV}ar(\cG)]$. 
\end{enumerate}
\end{proof}

\begin{proposition}
\label{prop-completeness}
For a program $\Pi$ and a goal $\leftarrow \cQ$, if there exists a derivation $\cD\equiv \tuple{(\leftarrow \cQ), id, \Pi} \hsld{*} \tuple{\square, \sigma, \Pi}$ then there exists a derivation $\cD'\equiv \tuple{\leftarrow \cQ', id, \Pi'} \sld{*} \tuple{\square,\sigma',\Pi'\cup\Pi_{reg}}$, where $\Pi'$ is the translation of each rule in $\Pi$,  $\cQ'$ is the goal translation of $\cQ$, $\Pi_{reg}$ is the set of rule registrations, that is, all the $reg$/3 assertions due to embedded implication solving, and $\sigma = \sigma'[{\cV}ar(\cG)]$.\QED{0pt}
\end{proposition}
\begin{proof}
By induction on the length of the derivation $\cD$ and Lemma~\ref{lem-step-completeness}.
\end{proof}

As mentioned above, Proposition~\ref{prop-completeness} constitutes a kind of completeness result. 
In the following we concentrate on the other direction, which leads to a sort of soundness result.

\newpage
\subsection{Proof of Part II}

\begin{proposition}
\label{prop-soundness}
Let  $\Pi'$ be the translation of a program $\Pi$ and $\cG'\equiv (\leftarrow \cQ'_1)$ the goal translation of a goal $\cG\equiv (\leftarrow \cQ_1)$, if there exists a derivation 
$\cD'\equiv \tuple{\leftarrow \cQ', \theta\cup\delta, \Pi'} \sld{*} \tuple{\square,(\theta\sigma)\cup\delta',\Pi'\cup\Pi_{reg}}$, 
where $\Pi_{reg}$ is the set of rule registrations, that is, all the $reg$/3 assertions due to embedded implication solving, 
then there exists a derivation 
$\cD\equiv \tuple{(\leftarrow \cQ), \theta, \Pi} \hsld{*} \tuple{\square, \theta\sigma, \Pi}$. 
The domains of the substitutions $\delta$ and $\delta'$ share variables with neither $\theta$ nor $\sigma$, and  $\theta\sigma = (\theta\sigma)\cup\delta'[{\cV}ar(\cG)]$.\QED{0pt}
\end{proposition}
\begin{proof}
The proof proceeds by induction on the length of the derivation $\cD'$.  Without loss of generality, thanks to the independence of the computation rule in the SLD operational mechanism, several steps in the derivation $\cD'$ can be  conveniently ordered. So, it is possible to group fragments of the derivation $\cD'$ in the translated program $\Pi'$, which correspond with the steps in the derivation $\cD$, in the program $\Pi$.

\begin{enumerate}
\item {\sf Base case($n=1$)}:  In this case the query $\cQ'\equiv p(\overline{s_n}, [], I, C, \epsilon)$ (translation of $\cQ\equiv p(\overline{s_n})$, since an initially launched goal is solved in the initial context $\epsilon$ and its list of shared variables is empty) and there must be a rule $p(\overline{t_n}, [], i, \epsilon, S^{C})$, with rule index $i$ (translation of the fact $p(\overline{t_n}) \in \Pi$), for which 
\[
\begin{array}{l}
  mgu(\{p(\overline{s_n}, [], I, C, \epsilon)= p(\overline{t_n}, [], i, \epsilon, S^{C})\})   \\
  ~~~~~~=mgu(\{\ol{s_n = t_n}, []=[], I=i, C=\epsilon, \epsilon=S^{C}\})   \\
  ~~~~~~= \sigma \{I/i, C/\epsilon, S^{C}/\epsilon\} =  \sigma \cup \delta_1   
\end{array}
\]
where $mgu(\{\ol{s_n = t_n}\})=\sigma$ must be solvable, $\delta_1 = \{I/i, C/\epsilon, S^{C}/\epsilon\}$, and its domain shares variables with neither $\theta$ nor $\sigma$. 
Therefore, the following one-step derivation $\cD$ is possible in the program $\Pi$: 
$\tuple{\leftarrow p(\overline{s_n}), \theta, \Pi} \hsld{} \tuple{\square, \theta\sigma, \Pi}$ .

\item {\sf Inductive case ($n > 1$)}: In the analysis of the inductive case we consider two possibilities, depending on whether the first step is performed with a sub-goal, which is an instance of the atom $\Rightarrow (I^{H},[X'], G', I^C, S^C)$, or not. This kind of sub-goal comes from embedded implications that appear in the body of some rule, or are submitted directly in the initial query proposed to the system.

\begin{enumerate}
\item First, consider that $\cQ'\equiv \leftarrow p(\overline{s_n}, [], I, C, epsilon)\land\cQ'_1$ (translation of $\cQ\equiv \leftarrow p(\overline{s_n})\land\cQ_1$) and the first step is performed with a rule $(p(\overline{t_n}, [ ], i, \epsilon, S^{C}) \leftarrow \cB')\in \Pi'$, with index rule $i$ (translation of $(p(\overline{t_n}) \leftarrow \cB)\in \Pi$),\footnote{
Note that in the translated program $\Pi'$, we can find rules of the form 
$p(\overline{t_n}, [\overline{X'}], I^{H}, S^{H}, S^C) \leftarrow reg(I^H, [\overline{X'}], C^R) \land chk(C^R, S^C) \land \cB'$ 
coming from the translation of the embedded implications in the body of the rules in $\Pi$ (see Definition 
3.5). However, this type of rule does not contribute to the first step of a derivation.
}
whose head unifies with the selected sub-goal: 
\[
\begin{array}{l}
  mgu(\{p(\overline{s_n}, [], I, C, epsilon)= p(\overline{t_n}, [], i, \epsilon, S^{C})\})   \\
  ~~~~~~=mgu(\{\ol{s_n = t_n}, []=[], I=i, C=\epsilon, \epsilon=S^{C}\})   \\
  ~~~~~~= \sigma \{I/i, C/\epsilon, S^{C}/\epsilon\} =  \sigma_1 \cup \delta_1   
\end{array}
\]
where $mgu(\{\ol{s_n = t_n}\})=\sigma_1$ must be solvable, $\delta_1 = \{I/i, C/\epsilon, S^{C}/\epsilon\}$ and its domain shares variables neither with $\theta$ nor with $\sigma$. 
Hence, derivation $\cD'$ proceeds thus:
\[
\begin{array}{l}
\tuple{\leftarrow p(\overline{s_n})\land\cQ'_1, \theta\cup\delta, \Pi'} \\
~~~~~~\sld{} \tuple{\leftarrow (\cB'\land\cQ'_1)(\sigma_1\cup\delta_1), (\theta\cup\delta)(\sigma_1\cup\delta_1), \Pi'}\\ 
~~~~~~\sld{*} \tuple{\square,(\theta\cup\delta)(\sigma_1\cup\delta_1)(\sigma_2\cup\delta_2),\Pi'\cup\Pi_{reg}}
\end{array}
\]

Now, because $mgu(\{p(\overline{s_n})= p(\overline{t_n})\})=mgu(\{\ol{s_n = t_n}\})=\sigma_1$ is solvable, there exists the step in $\Pi$:
$$\tuple{\leftarrow p(\overline{s_n})\land\cQ_1, \theta, \Pi}\hsld{} \tuple{\leftarrow (\cB'\land\cQ'_1)\sigma_1, \theta\sigma_1, \Pi}$$
on the other hand, by the inductive hypothesis, there exists a derivation in $\Pi$ such that:
$$\tuple{\leftarrow (\cB\land\cQ_1)\sigma_1, \theta\sigma_1, \Pi}\hsld{*} \tuple{\square,\theta\sigma_1\sigma_2,\Pi}$$
and the derivation $\cD$ in $\Pi$ can be constructed.

\item In this case, the first step is performed on a solving implication clause, that is $\cQ'\equiv \leftarrow (\Rightarrow (i, [\overline{X}], G', I^C, S^C)\land\cQ'_1)$ which is the translation of $\cQ\equiv \leftarrow (H \Rightarrow G)\land\cQ_1$. 
For simplicity, we will assume that $G'\equiv q(\overline{s_n})$ and the goal translation $G\equiv q(\overline{s_n}, L, I, C, S^C.I^C)$. Then the shape of the derivation $\cD'$ is:
{\small
\[
\begin{array}{l}
    \tuple{\leftarrow (\Rightarrow (i, [\overline{X}], G', I^C, S^C)\land\cQ'_1), \theta, \Pi'}  \\
    \sld{} \tuple{\leftarrow (get\_ci(I^C)\land reg\_rule(i, [\overline{X}], I^C, S^C)\land call(G')\land\cQ'_1), \theta, \Pi'}\\
    \sld{} \tuple{\leftarrow (reg\_rule(i, [\overline{X}], j, S^C)\land call(G')\land\cQ'_1)\{I^C/j\}, \theta\cup\{I^C/j\}, \Pi'}\\
     \sld{} \tuple{\leftarrow (assertz(reg(i, [\overline{X}], S^C))\land call(G')\land\cQ'_1)\{I^C/j\}, \theta\cup\{I^C/j\}, \Pi'} \\   
    \sld{} \tuple{\leftarrow (call(G')\land\cQ'_1)\{I^C/j\}, \theta\cup\{I^C/j\}, \Pi'\cup\{reg(i, [\overline{X}], S^C)\}} \\   
    \sld{} \tuple{\leftarrow (G'\land\cQ'_1)\{I^C/1\}, \theta\cup\{I^C/1\}, \Pi'\cup\{reg(0, [\overline{X}], S^C)\}} \\ 
    \sld{+} \tuple{\leftarrow Q'_1(\sigma_1\cup\{I^C/1\}\cup\delta_1), (\theta\sigma_1)\cup\{I^C/j\}\cup\delta_1,\Pi'\cup\{reg(i, [\overline{X}], S^C)\}\cup\Pi_{reg1}}  \\
    \sld{*} \tuple{\square, (\theta\sigma_1\sigma_2)\cup\{I^C/j\}\cup\delta_1\cup\delta_2,\Pi'\cup\{reg(i, [\overline{X}], S^C)\}\cup\Pi_{reg1}\cup\Pi_{reg2}}  
\end{array}
\]
}
with two clear parts. The first part corresponds to the HSLD step performed on $(H \Rightarrow G)$ using Rule 2 of Definition 
3.1. It groups the associated derivations submitted by the occurrence of embedded implications and their successive program contexts. Then there must exist the HSLD step  
$\tuple{\leftarrow (H \Rightarrow G)\land\cQ_1), \theta, \Pi} \hsld{} \tuple{Q_1\sigma_1, \theta\sigma_1,\Pi}$ in the program $\Pi$. 
As for the second part, by the inductive hypothesis, there must exist the HSLD derivation
$\tuple{Q_1\sigma_1, \theta\sigma_1,\Pi}  \sld{*} \tuple{\square, \theta\sigma_1\sigma_2,\Pi}.$ 
Combining both, the former step and the last derivation, we obtain the derivation $\cD$.
\end{enumerate}

\end{enumerate}
\end{proof}

\begin{corollary}
\label{cor-completeness}
Let  $\Pi'$ be the translation of a program $\Pi$ and $\cG'\equiv (\leftarrow \cQ'_1)$ the goal translation of a goal $\cG\equiv (\leftarrow \cQ_1)$, if there exists a derivation 
$\cD'\equiv \tuple{\leftarrow \cQ', id, \Pi'} \sld{*} \tuple{\square,\sigma',\Pi'\cup\Pi_{reg}}$, 
where $\Pi_{reg}$ is the set of rule registrations, that is, all the $reg$/3 assertions due to embedded implication solving, 
then there exists a derivation 
$\cD\equiv \tuple{(\leftarrow \cQ), id, \Pi} \hsld{*} \tuple{\square, \sigma, \Pi}$, and  $\sigma = \sigma'[{\cV}ar(\cG)]$. 
\QED{0pt}
\end{corollary}

\clearpage
\section{Related Work}
\label{app-related}

The term "hypothetical reasoning" appears in many important contexts in the philosophical and scientific literature.
Auguste Comte, founder of Positivism, is one of the first thinkers to highlight the importance of hypotheses in science \cite{Bou2020}. Although Comte does not establish laws for hypothetical reasoning, he begins the path, influencing (according to Michel Bourdeau \cite{Bou2014}) other thinkers such as Peirce and its abductive reasoning. According to that American philosopher, human thought has three modes of reasoning: deductive, inductive and abductive. "Abduction is the first step of scientific reasoning", because, as he says, "abduction is the process of forming explanatory hypotheses. It is the only logical operation that introduces a new idea" \cite{Dou2021}. 

However, our work is centered in a more specific area with a long tradition in the field of Logic Programming, in which the purpose is prospective: to propose hypotheses in order to evaluate its consequences.
Mainly, our work is influenced by those of Gabbay \cite{GR84,Gab85} and Bonner \cite{Bonner88,Bonner90,bonner1994hypothetical,Bonner97} and, to a lesser extent, by those of L.T. McCarty \cite{McCarty88a,McCarty88b}. 

Gabbay first deals with hypothetical implications in logic programming. In \cite{GR84}, {\color{black}they focused only on} addition operations because deletion is problematic{\color{black}; thus they} let it for another paper. Addition is essentially monotonic and deletion is not. We use a similar technique {\color{black}to} the one followed by Gabbay for implementing hypothetical implications, by asserting the antecedent to the program database, trying to derive the consequent and finally retracting the antecedent.
\cite{Gab85} investigates the logical properties of N-PROLOG and the way it relates to classical logic and the classical quantifiers. He also introduced negation as failure into N-PROLOG. He saw that success in the N-PROLOG computation of a goal $G$ from the database $P$ means logically that $P \vdash G$ in intuitionistic logic. 
It is credited that was Gabbay the first one to realize the important connection between hypothetical reasoning and intuitionistic logic \cite{bonner1994hypothetical}.

In \cite{McCarty88a}, he presents a clausal language that extends Horn-clause logic by adding negations and embedded implications (i.e., hypothetical implications --he was the one who first used this designation--) to the right-hand side of a rule, and interpreting these new rules intuitionistically in a set of partial models. Lately, in \cite{McCarty88b}, he shown that clausal intuitionistic logic has a tableau proof procedure that generalizes Horn-clause refutation proofs and it is proved  sound and complete.  

As it has been said, Bonner has extensive experience in this field, starting from a language with embedded implications (close to ours) and exploring its applications and formal properties, including results on complexity \cite{Bonner88,Bonner90,bonner1994hypothetical}. In his latest work on this topic \cite{Bonner97}, he broke with his initial works and he developed a logic programming language with a dedicated syntax in which users can create hypotheses and draw inferences from them. He provides two specific operations with a modal-like notation: hypothetical insertion of facts into a database ($Q[add: A]$ meaning that ``Q would be true if A were added to the database''), that has a well-established logic (intuitionistic logic) and hypothetical deletion ($Q[del: A]$ meaning that ``Q would be true if A were deleted from the database''). In this paper, he develop a logical semantics for hypothetical insertions and deletions (including a proof theory, model theory, and fixpoint theory). He analyses the expressibility of the language and he  shows that classical logic cannot express some simple hypothetical queries. However, we believe that the language introduced, with specific insertion and deletion operations, may have limitations compared to the one we have proposed in this work (e.g., only atoms can be inserted or deleted). Finally, he augmented the logic with negation-as-failure so that nonmonotonic queries can be expressed, a subject that we let as future work. 

Finally, another piece of related work is $\lambda$-Prolog, which uses a syntax based on the so-called ``hereditary Harrop formulas{\color{black}."} 
Thanks to this type of formulas, $\lambda$-Prolog subsumes a set of increasingly complex sublanguages, ranging from Horn clauses and higher-order Horn clauses to hereditary Harrop formulas.  This type of formula, for example, allows rules with bodies that contain (hypothetical) implications whose hypotheses are in turn rules.

The use of higher-order Horn clauses and a non-deterministic goal-directed search-based operational semantics (which is complete with respect to an intuitionistic sequent calculus) \cite{MNPS91} allows $\lambda$-Prolog the ability to perform hypothetical reasoning.  In practical and informal terms, the $\lambda$-Prolog operational mechanism performs operations similar to those performed by {\color{black}\textsf{Comp}} to solve a hypothetical implication ($H \Rightarrow G$): Assert $H$ to the rules of the program (creating a new context) and launch the goal $G$; if $G$ is successful, then ($H \Rightarrow G$) is also successful.  It is nothing other than the rule "AUGMENT" ($P \vdash (D \Rightarrow G)$ only if $P + {D} \vdash G$.), one of the operational rules that models the computation-as-goal-directed-search of $\lambda$-Prolog.

Therefore, both mechanisms are comparable, except that $\lambda$-Prolog fits into a more general and ambitious framework, while {\color{black}\textsf{Comp}} simply tries to extend the language of Horn clauses and the resolution principle with additional features, among which the hypothetical reasoning is found{\color{black}, as a platform for a fuzzy logic programming system}. 

What specific expressive capabilities for hypothetical reasoning $\lambda$-Prolog incorporates depends on the implementation.  For example, we know in the words of D. Miller himself\footnote{\url{stackoverflow.com/questions/65176668/?prolog-rejecting-hypothetical-reasoning-queries}} about Teyjus, an implementation of $\lambda$-Prolog, that "Teyjus does not permit implications to be used in top-level goals.  This is a characteristic that may change in the future when the compilation model is extended also to these goals but, for now, it means that some of the examples presented, eg, in Section 3.2, cannot be run directly using this system.''
Instead, the future implementation of {\color{black}\textsf{Comp}} is planned to be able to allow this by compiling the goal in the context of the loaded program before submitting it.

Moreover, $\lambda$-Prolog does not work with "negative assumptions", a matter that we have let marked as future work, and that we will undertake by following some ideas already proposed for the implementation of a Fuzzy Datalog system [Julian-Iranzo and Saenz-Perez, 2018].

Despite all the above and the possible relations between the foundations of our proposal and $\lambda$-Prolog, as we have just commented in this section, our work has its roots in the previous work carried out by Gabbay and Bonner.  On the other hand, the main contribution of this article is the development of efficient high-level implementation techniques for implementing a fuzzy logic programming system ({\color{black}\HBPLa}) with the possibility of hypothetical reasoning {\color{black} based on \BPLa\ \cite{RJ14JIFS,JR17FSS}}.  In such a system, assumptions imply compilations which with our proposal are possible to perform at compile-time.

}

\supp{
\bibliographystyle{acmtrans}

}
{}

\label{lastpage}
\end{document}